\documentclass[conference,compsoc]{IEEEtran}
\IEEEoverridecommandlockouts
\pdfoutput=1

\DeclareRobustCommand*{\IEEEauthorrefmarknum}[1]{%
    \raisebox{0pt}[0pt][0pt]{\textsuperscript{\footnotesize\ensuremath{#1}}}}

\usepackage{tikz}
\usepackage{amsmath}
\usepackage{amsthm}
\usepackage{caption}
\usepackage{multirow}
\usepackage[T1]{fontenc}  
\usepackage{amsfonts}  
\usepackage[numbers,sort&compress]{natbib}

\usepackage[caption=false,font=footnotesize]{subfig} 
\usepackage{booktabs}
\usepackage{algorithm}
\usepackage{algorithmic}
\usepackage{float}
\usepackage{enumitem}
\usepackage[breaklinks,colorlinks]{hyperref}

\usepackage[capitalize]{cleveref}
\Crefname{section}{Sec.}{Secs.}
\crefname{section}{Section}{Sections}
\Crefname{table}{Tab.}{Tabs.}
\crefname{table}{Table}{Tables}
\Crefname{figure}{Fig.}{Figs.}
\crefname{figure}{Figure}{Figures}

\begin{document}

\date{}

\title{SoK: On the Role and Future of AIGC Watermarking in the Era of Gen-AI}


\author{
    \IEEEauthorblockN{Kui Ren\IEEEauthorrefmarknum{1}\textsuperscript{,}\IEEEauthorrefmarknum{2}, Ziqi Yang\IEEEauthorrefmarknum{1}\textsuperscript{,}\IEEEauthorrefmarknum{2}, Li Lu\IEEEauthorrefmarknum{1}\textsuperscript{,}\IEEEauthorrefmarknum{2}, Jian Liu\IEEEauthorrefmarknum{1}\textsuperscript{,}\IEEEauthorrefmarknum{2}, Yiming Li\IEEEauthorrefmarknum{1}\textsuperscript{,}\IEEEauthorrefmarknum{2}\textsuperscript{,}\IEEEauthorrefmarknum{3}\\
    Jie Wan\IEEEauthorrefmarknum{1}\textsuperscript{,}\IEEEauthorrefmarknum{2}, Xiaodi Zhao\IEEEauthorrefmarknum{1}\textsuperscript{,}\IEEEauthorrefmarknum{2}, Xianheng Feng\IEEEauthorrefmarknum{1}\textsuperscript{,}\IEEEauthorrefmarknum{2}, Shuo Shao\IEEEauthorrefmarknum{1}\textsuperscript{,}\IEEEauthorrefmarknum{2}}
    \IEEEauthorblockA{\IEEEauthorrefmarknum{1} The State Key Laboratory of Blockchain and Data Security, Zhejiang University}
    \IEEEauthorblockA{\IEEEauthorrefmarknum{2} Hangzhou High-Tech Zone (Binjiang) Institute of Blockchain and Data Security}
    \IEEEauthorblockA{\IEEEauthorrefmarknum{3} College of Computing and Data Science, Nanyang Technological University}
    \IEEEauthorblockA{\{kuiren, yangziqi, li.lu, liujian2411, wanjie, yufen9, fengxianheng, shaoshuo\_ss\}@zju.edu.cn; liyiming.tech@gmail.com}
}

\maketitle

\begin{abstract}
The rapid advancement of AI technology, particularly in generating AI-generated content (AIGC), has transformed numerous fields, e.g., art video generation, but also brings new risks, including the misuse of AI for misinformation and intellectual property theft. 
To address these concerns, AIGC watermarks offer an effective solution to mitigate malicious activities. 
However, existing watermarking surveys focus more on traditional watermarks, overlooking AIGC-specific challenges.
In this work, we propose a systematic investigation into AIGC watermarking and provide the first formal definition of AIGC watermarking.
Different from previous surveys, we provide a taxonomy based on the core properties of the watermark which are summarized through comprehensive literature from various AIGC modalities. 
Derived from the properties, we discuss the functionality and security threats of AIGC watermarking. 
In the end, we thoroughly investigate the AIGC governance of different countries and practitioners. 
We believe this taxonomy better aligns with the practical demands for watermarking in the era of GenAI, thus providing a clearer summary of existing work and uncovering potential future research directions for the community.
\end{abstract}

\section{Introduction}

The rapid development of artificial intelligence (AI), especially large generative AI (GenAI) (e.g., ChatGPT \cite{achiam2023gpt} and Stable Diffusion \cite{rombach2022high}), has largely changed society and lifestyles in recent years. With the ability to generate precise and high-quality AI Generated Content (AIGC), GenAI can assist users in accomplishing various tasks in the real world such as literature reviewing \cite{touvron2023llama}, medical diagnosis \cite{thirunavukarasu2023large}, and painting \cite{xiao2024omnigen}. This significantly increases the efficiency and quality of human work and creation. 

However, the emergence of GenAI and AIGC not only offers new possibilities for creation but also brings a series of challenges and issues \cite{pang2024attacking,jovanovic2024watermark}.
Malicious users may adopt the AI model to generate harmful content that may conduct a series of malicious behaviors, including creating and spreading misinformation \cite{zhang2023watermarks,jiang2023evading}, stealing intellectual property \cite{zhang2024v2a,zhang2024editguard}, etc.
To address these issues, watermarking is regarded as a promising solution to prevent the abuse of AIGC, due to its properties, e.g., robustness and traceability.
Watermarking methods embed distinctive information into the watermarked data to verify authenticity, protect the copyright, or identify the creator.
Currently, most of the previous literature reviews primarily focus on traditional watermarking \cite{liu2024survey, hosny2024digital, zhong2023brief}. However, AIGC watermarking presents a more intricate challenge from a modality perspective, as the input and output of large GenAI may involve cross-modalities in contrast to traditional models which are centered around a single modality. Consequently, there remains a lack of comprehensive understanding for the security community regarding the challenges of AIGC watermarking and the taxonomy of existing works.

\begin{figure}[!tp]
    \centering
    \includegraphics[width=0.95\linewidth]{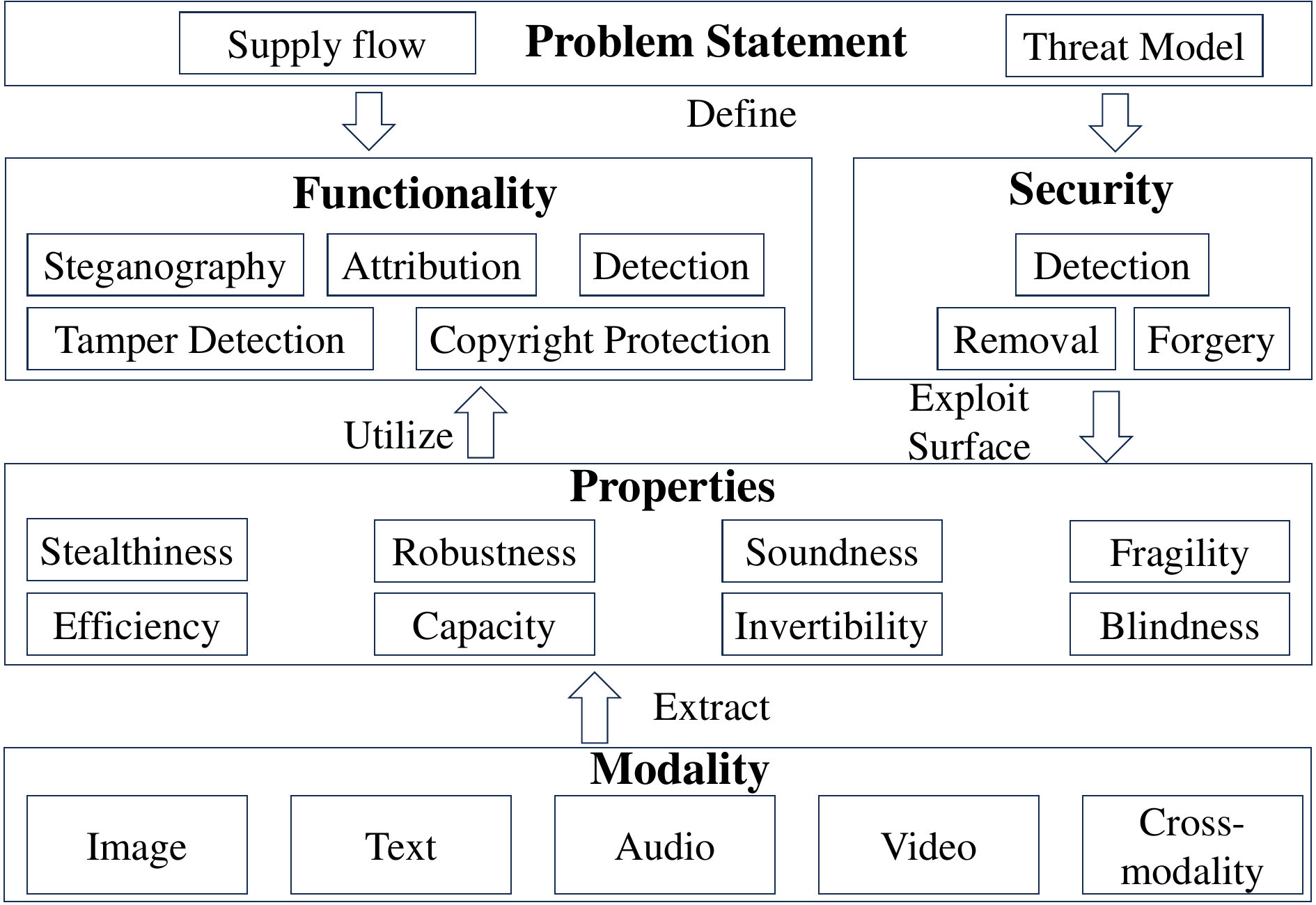}
    \caption{The overview of this paper.}
    \label{fig:overview}
    \vspace{-1em}
\end{figure}

In this work, we propose a comprehensive investigation into watermarking in the era of GenAI and provide the first formal definition.
To align with the practical demands of AIGC watermarking, we start our study with the literature from the perspective of AIGC modality, and based on this, we summarize the core properties of AIGC watermarks. We believe this is helpful for practitioners to have a sense of the capability boundary of watermarking in AIGC applications. 
Derived from these properties, we discuss the functionality and security issues in the application of the AIGC watermark.
From this taxonomy, we can identify potential application scenarios and possible vulnerabilities of the AIGC watermark and encourage future work for the community.
Finally, we provide a thorough discussion of AIGC governance from the perspectives of countries and practitioners. 
In \cref{fig:overview}, we provide the outline of this work, which shows the overview of the AIGC modality, the formal definition of the AIGC watermark, the functionality, and the security issues.

We provide the formal definition of the AIGC watermark in \cref{sec:problem:threat_model}, regulating the generation of the AIGC watermark as the flow-like supply chain.
Based on such a definition, we formalize the threat model to point to the potential attack node in the whole lifetime of the AIGC watermark.
In contrast to traditional watermarks, which are typically added externally, AIGC watermarks could be embedded directly within the supply chain, bringing unique qualities that improve security, flexibility, and cross-media adaptability.
We discuss the modalities of the AIGC watermark in \cref{sec:Modality}, with the challenges and techniques for embedding watermarks in AI-generated content across traditional modalities and the newly cross-modality 
brought by AIGC.
Based on the existing work in different modalities, we summarize the properties of AIGC watermarks in \cref{sec:properties} from the basic level and the advanced level. 

These key properties cater to the demands of AI-driven applications, prioritizing security and authenticity.
Spanned from the AIGC watermark properties, we conducted an in-depth analysis of the principles of the AIGC watermark in \cref{sec:Functionality} and the security issues in \cref{sec:security} that can be exploited in the supply chain.
We taxonomize the applications of the AIGC watermark based on the intended use, e.g., copyright protection, steganography, and tamper detection, where each application may relate to one or more properties of the AIGC watermark.
We introduce the security issues of the AIGC watermark and highlight the vulnerabilities that adversaries can exploit.
We motivate the community to explore the functionality and the threat of the watermark on the unexplored combination of properties.
Additionally, we summarize the existing regulations on the watermark in different regions and countries in \cref{sec:Regulations}.

We hope that our systematization can act as a milestone toward the AIGC watermark. 
Most importantly, we provide guidelines to encourage the communities to explore more potential applications and mitigate more security issues in the application of the AIGC watermark.

In summary, we make the following contributions:
\begin{itemize}[leftmargin=*]
    \item \textbf{Formal definition of AIGC watermark:} To the best of our knowledge, we provide the first formal definition of the AIGC watermark, combining the phase of embedding and the modalities in the large AI models.
    \item \textbf{Review of properties of AIGC watermark:} We conduct a detailed review and analysis of existing research, extracting various attributes of AIGC watermarks, including some attributes inherent to traditional watermarks, as well as emerging attributes that characterize watermark performance in AIGC.
    \item \textbf{Systematization of the functionality and security of AIGC watermark:} We first summarize the functionality and security of AIGC watermarking based on the properties, to show how each existing attack method performs with the guideline under the properties.   
    \item \textbf{Potential directions of the future work:} We encourage the community to explore more on those properties of the AIGC watermark that have never been studied before and propose some potential directions.
\end{itemize}

\section{Problem Statement of AIGC Watermark}\label{sec:problem:threat_model}

The typical pipeline of AIGC watermarking involves five key elements: the dataset, the user, the generative model, the AIGC, and the watermark. 
The embedding process of the watermark can be operated with different elements, i.e., different stages of the generation of AIGC.

In this section, we define the watermark as a flow-like supply chain and provide the threat model to regulate the capability of the adversary and introduce the attack surface.

\begin{figure}[t]
    \centering
    \includegraphics[width=\linewidth]{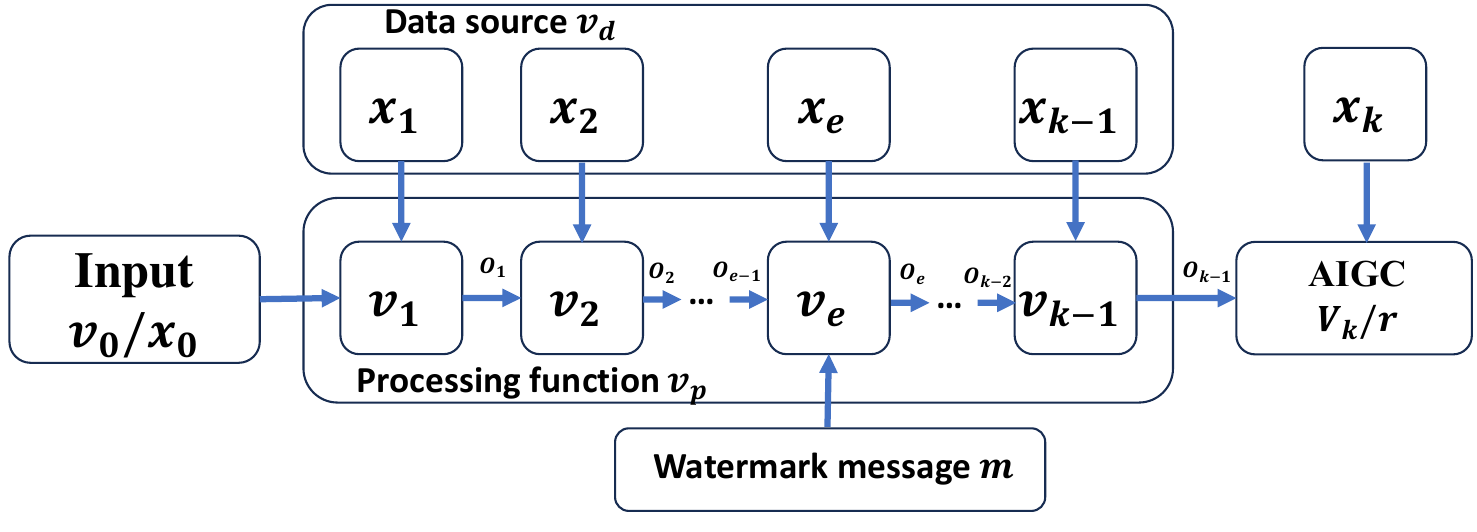}
    \caption{An illustration of the flow-like supply chain of AIGC watermarking, which starts from the composition of the input ($v_0$), and ends at the final output returned to the user. Each sub-point in the generation process of AIGC can be regarded as a node.}
    \label{fig:flow}
    \vspace{-0.5em}
\end{figure}

\subsection{Formal Definition of AIGC Watermark}\label{sec:Formal Definition}

AIGC watermark refers to watermarks extracted from and verified in AI-generated content, with the aim of mitigating the AIGC-posed challenges including generated content detection, model attribution, user identification, copyright protection, tamper detection, etc.

To define it formally, we denote the supply chain of AIGC as a rooted tree $T = (\mathbb{V}, \mathbb{E}, r)$, illustrated in \cref{fig:flow}, where $\mathbb{V} = V_d \cup V_p$ is the set of vertices, with leaf nodes $v_d\in V_d$ representing the data sources (e.g., input prompt and conditions) and internal nodes $v_p\in V_p$ representing the processing functions (e.g., one step of diffusion process). The input of $V_p$ is the output of its parent nodes. The root node $r$ outputs the final generated content. i.e., AIGC. 
The directed edges $\mathbb{E}$ represent the data flow between nodes, pointing towards the root.

\begin{table}[t]
    \centering
    \caption{Notations.}
    \begin{tabular}{cl}
        \toprule
        \textbf{Symbol} & \textbf{Description} \\
        \midrule
        $T=(\mathbb{V},\mathbb{E},r)$ & Rooted tree representing AIGC supply chain \\
        $V_d$ & Set of leaf nodes (data sources) \\
        $V_p$ & Set of internal nodes (processing functions) \\
        $r$ & Root node outputting final AIGC \\
        $P$ & Leaf-to-root path $(v_0,v_1,...,v_k)$ \\
        $v_e$ & Node where the watermark is introduced \\
        $S$ & Set of subsequent nodes after $v_e$ \\
        $x$ & Initial input \\
        $x_i$ & External input to node $v_i$ \\
        $o_i$ & Output of node $v_i$ \\
        $y$ & Final output without watermark \\
        $y'$ & Final output with watermark \\
        $m$ & User-specific watermark message \\
        $m'$ & Extracted watermark message \\
        $E$ & Watermark embedding function \\
        $D$ & Watermark extractor function \\
        $D_m$ & Distance function for messages \\
        $D_y$ & Distance function for outputs \\
        $\mathcal{W}$ & Watermarking module \\
        $\mathbb{F}$ & Set of possible watermarking functions \\
        $\mathbb{A}$ & Set of possible attack vectors \\
        $\epsilon$ & Maximum allowable watermark magnitude \\
        \bottomrule
    \end{tabular}
    \label{tab:notation}
    \vspace{-1em}
\end{table}

A watermarking method first identifies a leaf-to-root path \( P = (v_0, v_1, v_2, \ldots, v_k)\) in the tree \( T \) where $v_k=r$ such that watermarking on a particular node \( v_e \in P \) can affect the subsequent nodes in the path \( S = \{ v_j\in P\ |\ e < j \leq k\} \)\footnote{
    We note that the methodology of watermarking is akin to that of supply chain attack \cite{2024Supply}, although the former serves a positive purpose while the latter is a negative one.
}.
We denote the initial input $v_0$ as $x$, input data of $v_i$ from nodes not in the path $\{v |(v, v_i)\in \mathbb{E}, v\neq v_{i-1}\}$ as $x_i$, the final output (without watermark) as $y$ and the propagation operation from the last node to the next node in the flow as $\circ$.
Thus, we have:
\begin{align}
y &= (v_k(x_k,\cdot) \circ v_{k-1}(x_{k-1},\cdot) \circ ... \circ v_1(x_1,\cdot))(x) \\
&= T(x)
\end{align}
For watermarking, $v_e$ is replaced with a function $v'_e$ which accepts another input, i.e., the user-specific message $m \in \{0,1\}^{d_m}$. Thus, we have:
\begin{align}
y' &= (v_k(x_k,\cdot) \circ ... \circ v'_e(x_e,\cdot, m) \circ ... \circ v_1(x_1,\cdot))(x)\\
&= T'(x,m)
\end{align}
An extractor function $E$ extracts the message $m'$ from the last output:
\begin{align}
m' = E(y')
\end{align}
The message consistency before and after the process should be maximized, and the utility changes in the output due to watermarking should be minimized. 
Thus, given two distance functions \(D_m\) (e.g., binary cross-entropy loss) and \(D_y\) (e.g., perceptual distance of images) respectively, and let \(\mathbb{F}\) be the set of all possible watermarking functions and introduce $\lambda$ to control the relative weight between the two distance functions.
The primary objective is:
\begin{align}
\min_{v'_e\in \mathbb{F}} \quad D_m(m, m') + \lambda D_y(y, y')
\end{align}
The watermarks are extracted from and verified in the generated content $y$, i.e., the AIGC watermark\footnote{
    Watermarks verified within other entities, such as model weights for intellectual property protection, fall outside the scope of this paper.
}.

\subsection{Threat Model in AIGC Watermark}

\subsubsection{Attack Surface}

We first assume the attacker fully controls the target model, to provide a comprehensive overview of all the possible attack surfaces. 
To attack the AIGC watermark, there are two typical objectives:

\noindent\textbf{Untargeted Attack (Removal).} The objective is to remove the watermarks without sacrificing the utility:
\begin{align}
    \max_{A\in\mathbb{A}} \quad D_m(m,m_A') \quad s.t. \quad D_y(y, y_A') \leq \epsilon.
\end{align}

\noindent\textbf{Targeted Attack (Overwrite/Forgery).} The objective is to overwrite or forge the watermarks with a target message without sacrificing the utility:
\begin{align}
    \min_{A\in\mathbb{A}} \quad D_m(m_{t},m_A') \quad s.t. \quad D_y(y, y_A') \leq \epsilon,
\end{align}
where $A$ denotes an attack vector, $m_A'$ and $y_A'$ denote extracted message and the content after attacking, $\mathbb{A}$ denotes the set of all possible attack vectors, $m_{t}$ denotes the target message, and $\epsilon$ is the maximum allowable watermark magnitude.

The propagation of the watermark can be intercepted at any subsequent phase.
Thus, for each subsequent node $v_i\in S$, there are three attack surfaces:
\begin{itemize}[leftmargin=*]
    \item Manipulating $x_i$, 
        e.g., modify parameters of the generative model.
    \item Manipulating $v_i$, 
        e.g., modify structures of the generative model.
    \item Manipulating the intermediate output $o_i$, 
        e.g., transform a generated image via JPEG compression.
\end{itemize}

Moreover, two additional attack surfaces exist in the watermarking scheme itself, which include manipulating $v'_e$ and manipulating $E$.
For instance, these could be done via cybersecurity attacks including malware and social engineering attacks.
To sum up, given $|S|=k-e$ subsequent nodes after watermarking, there are $3(k-e)+2$ attack surfaces.

\subsubsection{Attacker's Capability}
However, in practice, the attacker's capabilities are constrained at different levels, resulting in different attack settings.

\begin{itemize}[leftmargin=*]
    \item White-box setting. The adversary has full access to the victim model and the watermarking algorithm, which performs a more powerful analysis of the generation or evaluation process of the AIGC watermark.
    \item Black-box setting. The adversary has only access to the input-output pair, e.g., API access to the victim model.
    \item Gray-box setting. The adversary can manipulate input data, apply common transformations (e.g., compression or filtering), and access both the watermarked and original data. While they cannot fully control or reverse-engineer the embedding process, their partial knowledge allows them to weaken or attempt to remove the watermark, testing the system's robustness under limited attack conditions.
\end{itemize}

\section{AIGC Watermarking from Modality Perspective}\label{sec:Modality}
Watermarking techniques vary with the modality of the AI-generated content. In this section, we investigate the features of different modalities and the shortcomings of watermarking these modalities when generated by AI compared with traditional watermarking methods. 
Then we summarize widely-studied methods of watermarking each modality. 
Finally, we discuss the remaining challenges of watermarking in the era of GenAI across different modalities.

\subsection{Text}
Text data is highly discrete compared with other media, typically using words or tokens as basic units with high information density.
As a result, modification of just a few words can significantly alter the semantics of the text. 
Due to these features, traditional text watermarking techniques often embed watermarks by altering the format \cite{format1,format2,format3}, syntactic or semantic \cite{yang2022tracing,lexcialsubstitution2,lexcialsubstitution3} of text. 
However, these watermarks either struggle to withstand format-based changes such as copy-pasting and retyping or fail to preserve the original semantic information of the text.

For AI-generated text, researchers propose to embed watermarks by making modifications to the probability distribution of tokens during text generation rather than directly changing the text, reducing the impact on the text's semantics. According to the way used to modify the distribution, text watermarking techniques can mainly be divided into three categories: logits-based, sampling-based, and training-based methods. The logits-based methods \cite{sir,kirchenbauer2023watermark,unforgeable,zhao2023provable} embed hidden patterns into text by directly modifying the logit values of each token during the next token prediction and thus influencing their probabilities to be sampled. While sampling-based methods \cite{kth,aar} embed watermark by slightly modifying the sampling method of LLM, indirectly influencing the sampling probabilities of tokens and having less impact on text quality compared to the logits-based method. 
The training-based algorithm fine-tunes the model with watermarked text data to learn the hidden pattern in watermarked text \cite{learnability} or uses reinforcement learning for LLM finetuning so that the generated text can be detected by a specific watermark detectors \cite{reinforcement}. 

Despite the existence of text watermarking algorithms above, embedding watermarks in AI-generated text remains a challenge, particularly in low-entropy scenarios where it is hard to embed the watermark without greatly changing text's semantics and publicly verifiable scenarios where the watermark needs to remain unforgeable while watermark detectors should be publicly available to users.

\subsection{Image}
Digital images possess distinct features that make them suitable for watermarking.
For instance, an image comprises numerous pixels, providing an expansive space where watermark data can be embedded.

According to the domain of watermarks to be embedded, traditional image watermarking techniques can be mainly divided into spatial-based and frequency-based methods. The spatial-based methods~\cite{kalker1999video,chopra2012lsb} embed watermark by modifying the image's pixel or bit stream while frequency-based methods~\cite{ernawan2021improved,mohammed2023blind} achieve watermark embedding by leveraging frequency transformation algorithms. However, they are either vulnerable to spatial attacks or bring significant computation overhead with transformation operations, making them inefficient and hard to apply in AIGC.

In the era of GenAI, watermarking methods for images can be classified into training-free methods and training-dependent methods, depending on whether additional training is required. 
Training-free methods~\cite{wen2023tree,yang2024gaussian} inject specific patterns into the noise of diffusion model during image generation for watermark embedding. 
In the detection stage, the noise can be recovered utilizing an inversion algorithm and then the watermark is extracted. 
The training-dependent methods \cite{tan2024waterdiff,lukas2023ptw,zhao2023recipe,zeng2023securing,fernandez2023stable}
generally add an extra watermarking module within the AI generation pipeline or fine-tune the generative model parameters.

Although these methods have an insignificant impact on the model’s efficiency during image generation, they introduce additional computational burdens during the model training or watermark detection stage. Moreover, there remains a challenge in balancing the impact of image quality and the robustness of the watermark.

\subsection{Audio}
Audio is characterized by its time-frequency properties, waveform amplitude, phase information, and complex spectral structures while it is commonly subjected to noise addition and format conversion, which can easily distort or erase information.
Due to these features, audio watermarking requires maintaining high fidelity while ensuring robustness under various transformations.

Similar to image watermarking, traditional audio watermarking techniques can also embed watermarks by modifying audio in various domains, achieved by post-processing via transformations or deep neural networks~\cite{chen2023wavmark,liu2023dear}. 
Although traditional audio watermarking techniques are relatively mature, they still face challenges in terms of security and efficiency, particularly in AIGC scenarios.

For AI-generated audio, mainstream audio watermarking techniques typically involve adding an extra watermark module to ensure the generated audio contains a watermark. For example, ~\cite{cheng2024hifi} pretrains a watermark encoder and decoder, then fine-tunes the model with the watermark decoder involved to enable the generation of watermarked audio. ~\cite{zhou2024traceablespeech} use a neural network to extract features from the watermark message and fuse it into the audio generation process, with an extractor used to retrieve the watermark. Except for these methods, ~\cite{san2024latent} propose using pre-watermarked audio as training data, allowing the trained generative model to produce watermarked audio.

Although current watermarking methods have been able to generated watermarked audio with great quality. The robustness of these watermarks under evolving attacks such as audio recapturing remains a critical challenge.

\subsection{Video}
A video consists of a sequence of images, known as frames, that play in a specific order along a timeline.

Traditional watermarking techniques depend on manually crafted features, such as applying specific transforms~\cite{asikuzzaman2016robust} or using perceptual masking~\cite{campisi2005video}, or deep neural networks ~\cite{weng2019high,mishra2019vstegnet,luo2104dvmark,zhang2023novel,mou2023large} to encode watermarks to videos. However, they either remain vulnerable to various distortions~\cite{asikuzzaman2017overview} or add extra computation cost during watermark embedding, leading to a reduction in efficiency when they are applied to AIGC scenarios.

For AI-generated videos, there has been limited research exploring watermarking techniques specifically designed for video generation. Currently, mainstream video generation techniques generate video by creating images \cite{wen2023tree,yang2024gaussian} for each frame. Therefore, watermarking techniques applied to AI-generated images can also be used for embedding watermarks in videos. For example, DeepMind introduced SynthID \cite{synthid}, which supports directly embedding a watermark into the pixels of each video frame.

However, these methods not only lower the efficiency of video generation but also overlook correlations between frames. Thus, directly applying these methods may result in a decrease in the quality of the generated video and a shift in video content. Despite the rapid development of AI video generation techniques, watermarking algorithm especially for AI-generated videos remains an ongoing challenge.

\subsection{Cross-Modality}
Cross-modally refers to tasks that integrate and process information from multiple distinct data modalities, such as text, images, audio, and video while learning correlations between these modalities.

The watermarking techniques mentioned in the previous sections that involve multiple modalities typically embed watermarks utilizing the feature of a single modality \cite{wen2023tree,yang2024gaussian,synthid} and treating data from different modalities independently, which overlooks the correlations between modalities. As a result, the generated watermarked data may suffer from content mismatches and quality degradation. 

To solve this problem, cross-modal watermarking methods in the AIGC era take advantage of feature alignment across modalities by leveraging the powerful feature extraction capabilities of deep neural networks to enhance watermark robustness or reduce the impact of watermark on generated content. For instance, AVSecure \cite{guo2024avsecure} and $V^2A$-Mark \cite{zhang2024v2a} utilize both visual feature and audio features, enhancing the robustness of watermarks and their ability to tamper detection. \cite{liu2023t2iw} integrates both text and image features, thereby improving watermark robustness while minimizing the distortion of the image content.

While cross-modal watermarking takes advantage of the correlations between different modalities, it also faces the limitation of relying on a single modality. Effectively leveraging the benefits of cross-modal techniques while minimizing their drawbacks requires further investigation.

\section{Properties of AIGC Watermark} \label{sec:properties}
In contrast to traditional watermarking methods, which are commonly applied as an afterthought to content, AIGC (AI-Generated Content) watermarking is often embedded within the generative process itself. 
This intrinsic association differentiates AIGC watermarks by allowing for tighter integration and a greater level of security, as they are harder to remove without compromising content quality. 
Furthermore, AIGC watermarking can support diverse content types and modalities generated by advanced models, such as images, audio, and text, leveraging cross-modal embedding and providing enhanced flexibility and adaptability. 
This section provides a comprehensive investigation of the properties of the AIGC watermark, which guides our discussion in \cref{sec:Functionality} (Functionalities) and \cref{sec:security} (Security).

\subsection{Basic Properties of Traditional Watermark}
Traditional watermarking schemes, including those used in AIGC, are evaluated based on several key properties that ensure their effectiveness, security, and applicability in diverse use cases. 
These properties, which we summarize from eight key perspectives, form the foundational attributes that define the quality of a watermarking scheme. The following subsections provide an in-depth examination of these properties, including stealthiness \cite{aberna2024digital, begum2020digital, agarwal2019survey}, robustness \cite{aberna2024digital, wang2023data, begum2020digital}, soundness \cite{fairoze2023publicly, christ2024undetectable}, fragility \cite{wang2023data, asikuzzaman2017overview, langelaar2000watermarking}, efficiency \cite{begum2020digital, agarwal2019survey, anand2021watermarking}, capacity \cite{aberna2024digital, begum2020digital, agarwal2019survey}, invertibility \cite{wang2023data, begum2020digital}, blindness \cite{wang2023data, jiao2019review, wan2022comprehensive} and uniqueness \cite{ma2023generative}. They serve as essential metrics for assessing the performance of AIGC watermarks. 
While AIGC introduces new complexities, these basic properties remain critical in evaluating their effectiveness in protecting and identifying generated content.

\subsubsection{Stealthiness}  
Stealthiness \cite{aberna2024digital, singh2013survey}  refers to the ability of the watermark to remain perceptually indistinguishable from the original content, ensuring that the watermarked content appears identical to the original. 

Formally, let $x$ denote the original content and $y'$ the watermarked content; the scheme is deemed imperceptible if:
\begin{equation}
    {D_v}(x,{y'}) \leq {\epsilon}
\end{equation}
where $D_v$ is a perceptual similarity metric (e.g., Peak Signal-to-Noise Ratio, Structural Similarity Index). A lower $D_v$ score indicates higher imperceptibility.

\subsubsection{Robustness}
Robustness \cite{aberna2024digital, wang2023data} refers to the watermark's ability to persist despite various transformations or modifications applied to the watermarked content. 
Formally, given a watermarked content $y'$ and a set of allowable transformations $\mathbb{T}$ (see \cref{sec:security} for more details), a watermarking scheme is considered $\epsilon$-robust if
$ \forall t \in \mathbb{T}: D_m(m, E(t(y'))) \leq \epsilon $.
The robustness requirement can be quantified through the survival rate $\rho$ of the watermark:
\begin{equation}
    \rho = \frac{|\{T \in \mathbb{T} | D_m(m, E(T(y'))) \leq \epsilon\}|}{|\mathbb{T}|}.
\end{equation}
A higher $\rho$ indicates stronger robustness of the watermarking scheme against the specified set of transformations. However, there exists an inherent trade-off between robustness and other properties such as stealthiness and discriminability, necessitating careful balance in practical applications.

\subsubsection{Soundness} 
Soundness \cite{fairoze2023publicly, christ2024undetectable}, in the context of watermarking, refers to the property that an adversary cannot forge watermarked content without knowledge of the secret key. 
This ensures that genuinely watermarked content is distinct from any forged content.
A publicly-detectable watermarking scheme (PDWS) \cite{fairoze2023publicly} for an auto-regressive model over token vocabulary $\mathcal{T}$ consists of three algorithms: \textit{Setup}, \textit{Watermark}, and \textit{Detect}. The \textit{Setup} algorithm generates a public key pair $(\textit{sk}, \textit{pk})$ based on the security parameter $\lambda$. The \textit{Watermark} algorithm uses the secret key $\textit{sk}$ to generate a watermarked text $t$ based on a given prompt $\rho$. The \textit{Detect} algorithm verifies whether a given candidate text $t^*$ is watermarked, outputting \textit{true} if it is, or \textit{false} otherwise. 
A negligible function \texttt{negl($\lambda$)} decays faster than the inverse of any polynomial as $\lambda$ tends to infinity.

Let $E$ represent the event where an adversary, with access to the watermarking oracle $\text{Watermark}_{sk}(\cdot)$ and the public key $pk$, produces a watermarked content $t^*$ such that:

\begin{equation}
\left\{
\begin{array}{l}
\text{Detect}_{p_k}(t^*) = \text{true} \quad \land \\
\text{non\_overlapping}_{k}(t^*, t_1, t_2, \dots) = \text{true}
\end{array}
\right\}
\end{equation}

Here, the predicate $\text{non\_overlapping}_k(t^*, t_1, t_2, \ldots)$ ensures that $t^*$ does not share a $k
$-length window of tokens with any genuinely watermarked texts $t_1, t_2, \ldots$ The soundness of the scheme is defined such that:

\begin{equation}
\Pr\left[
\begin{array}{l}
(sk, pk) \leftarrow \text{Setup}(1^{\lambda}) \\
t^* \leftarrow \mathcal{A}^{\text{Watermark}_{sk}(\cdot)}(pk)
\end{array}
\right]
\left[ E \right] \leq \text{negl}(\lambda) 
\end{equation}

\subsubsection{Fragility} 
Fragility \cite{wang2023data, asikuzzaman2017overview, langelaar2000watermarking} in watermarking describes the sensitivity of the watermark to various alterations or attacks on the watermarked content, enabling verification of content integrity and detection of unauthorized modifications. 
\textit{Fragile} watermarks are particularly useful in contexts where authenticity and tamper detection are critical, as they reveal alterations by failing to withstand even minor changes to the host data. 
\textit{Semi-fragile} watermark allows for slight modifications to host data within a certain range, which can tolerate some non-malicious operations, such as compression, slight noise, and brightness adjustments, without becoming invalid due to these minor changes.

Fragility is inherently opposite to robustness, as fragile watermarks are not designed to survive typical attacks or transformations. The choice between fragility and robustness ultimately depends on the application’s primary goals. 
In environments where content integrity is critical, fragility is more advantageous as it reveals even minor alterations.

\subsubsection{Efficiency} 
Efficiency \cite{begum2020digital, agarwal2019survey, anand2021watermarking} in watermarking refers to the computational speed and resource requirements involved in embedding and extracting the watermark from the multimedia content. 

Efficiency is influenced by factors, e.g., the watermarking algorithm’s complexity, the data size of the host media, and the hardware used for processing. A high degree of efficiency may sometimes conflict with the robustness and fidelity of the watermark. For instance, complex algorithms may offer higher resistance to attacks but require more processing power and time, impacting efficiency. 
Conversely, simpler algorithms may lack robustness against sophisticated attacks. 
In summary, achieving optimal efficiency in watermarking requires a careful balance between computational demands and robustness, tailored to the application’s specific requirements.

\subsubsection{Capacity} 
Capacity \cite{aberna2024digital, begum2020digital, agarwal2019survey} in watermarking refers to the amount of information that can be embedded within the multimedia content, often quantified as bit-per-pixel in images or bit-per-frame in videos. Capacity is not only a measure of data quantity but also impacts the scheme’s overall functionality in applications requiring high distinguishability between multiple watermarked instances. 
For instance, the classical watermarking algorithm, Least Significant Bit (LSB) \cite{kalker1999video,chopra2012lsb} embeds data by altering the least significant bits of pixel values in an image, without causing noticeable distortion to the human eye.

It is important to balance capacity with other watermark properties, as increasing capacity often risks reducing the watermark’s stealthiness or robustness. 
Consequently, watermark capacity must be carefully optimized alongside other performance metrics to ensure the watermark remains unobtrusive and resilient in its intended application environment.

\subsubsection{Invertibility} 
Invertibility \cite{wang2023data, begum2020digital, mohanarathinam2020digital} in watermarking, also known as reversibility, refers to the capability of a watermarking scheme to allow the embedded watermark to be removed from the watermarked content, enabling complete recovery of the original content. Invertibility guarantees not only the watermark’s full extraction but also the exact reconstruction of the multimedia content. Formally, a watermarking scheme is defined as invertible if:
\begin{equation}
D(E(x,m))=(x,m)
\end{equation}
where $D(E(x, m))$ represents the result of fully extracting $m$ and restoring $x$ without any residual alterations. 

Although invertibility enhances authenticity and content integrity, it may present a trade-off with other properties such as robustness. 
If robustness is increased, achieving precise invertibility becomes more difficult, as the scheme must handle both extraction and full reconstruction of the original content after enduring potential distortions.

\subsubsection{Blindness} 
Blindness \cite{wang2023data, jiao2019review, wan2022comprehensive} in watermarking refers to how much information is required to extract the original data from the encoded media. Specifically, a blind watermarking scheme enables the extraction of the embedded watermark without requiring access to the original cover media, making it highly practical for real-world applications. Semi-blind techniques require only the original data, whereas Non-blind techniques require both the original cover media and data in order to perform data extraction.

It is necessary to detect and extract the message without the original media. Consequently, blindness not only increases the practicality of the watermarking scheme but also supports its reliability in diverse applications where the original media cannot feasibly be retained or accessed.

\subsubsection{Uniqueness}
Uniqueness \cite{ma2023generative} ensures each watermark is distinct to a specific content creator or user, allowing accurate identification without overlap. Uniqueness is measured by the detector’s ability to differentiate its own watermarked outputs from those of other users.
Formally, let $\mathcal{D}_i$ and $\mathcal{D}_j$ be detectors associated with subjects $i$ and $j$, respectively. A watermarking scheme achieves uniqueness if:
\begin{equation}
\Pr(\mathcal{D}_i(y'_i) = \text{true}) \gg \Pr(\mathcal{D}_j(y'_i) = \text{true}) \quad \forall i \neq j
\end{equation}
where $y'_i$ is the watermarked content uniquely assigned to subject $i$. This ensures that each detector $\mathcal{D}_i$ accurately identifies its own watermarked content, thus preventing cross-user detection or confusion and maintaining secure ownership attribution.

\subsection{Advanced Properties of AIGC Watermark}

AIGC watermarks are inherently designed to be embedded within the content generation process, providing a seamless and robust means of authenticating and tracing content. 
Unlike traditional watermarks that often serve as external add-ons, AIGC watermarks integrate directly into the generative pipeline, offering unique attributes that enhance security, flexibility, and cross-media applicability. 
Key properties of AIGC watermarking, including endogeneity, cross-modal attributes, plug-and-play capability, and accessibility, address the needs of AI-driven applications, where security, authenticity, and ease of verification are paramount.

\subsubsection{Endogeneity} \label{sec:properity_advanced_endogeneity}
Endogeneity in AIGC watermarking refers to the intrinsic relationship between the watermark and the supply chain itself. Unlike traditional watermarks that are typically added as a post-processing step, endogenous watermarks are inherently tied to the generation process. 
This tight integration enhances security by making watermarks harder to remove without degrading content quality, while also improving efficiency by eliminating separate watermarking steps.

Formally, let $T$ be a supply chain. Given data sources $X$, traditional watermarking can be expressed as:
\begin{equation}
    y' = \mathcal{W}(T(X))
\end{equation}
where $y'$ is the watermarked output, and the watermark is applied after generation.
In contrast, endogenous watermarking is integrated into the supply chain itself, which can be represented as:
\begin{equation}
    y' = T_w(X)
\end{equation}
where $T_w$ is the watermark-aware process that directly generates watermarked content. 

For instance, this integration can be achieved by modifying the parameters of a generative model $v_i$ in the supply chain.
Consequently, the watermarking function becomes an integral part of the model's parameters $\theta_w$:
\begin{equation}
    \theta_w = f(\theta, \mathcal{W})
\end{equation}
where $\theta$ represents the original model parameters and $f$ is a function that integrates the watermarking scheme into the model parameters.

\subsubsection{Cross-Modal Attributes}

Cross-modal attributes in AIGC watermarking refer to the capability of embedding watermarks across various content formats or media types generated by the model. 
This approach contrasts with traditional watermarks that are typically restricted to a single medium, enabling cohesive authentication across images, text, audio, and other content types within a unified framework.

This cross-modal consistency increases the difficulty of removing or altering the watermark, as it would require a multi-faceted adjustment across diverse formats, thus reinforcing overall security. The ability to sustain watermark integrity across media types not only simplifies the verification process but also enhances robustness against tampering, making it highly suitable for applications demanding seamless verification across varied content formats.

\subsubsection{Plug-and-Play Capability}

The plug-and-play capability in AIGC watermarking allows watermark embedding without retraining or fine-tuning the generative model. 
This is particularly valuable for large models, where retraining is computationally intensive. 
Plug-and-play frameworks integrate seamlessly into existing generative pipelines as auxiliary modules, thereby embedding watermarks while preserving the model’s original parameters.

Formally, let $v_i$ represent the generative model with parameters $\theta$. In plug-and-play watermarking, the output $y'$ is generated as:
\begin{equation}
y' = \mathcal{W}(v_i(x, \theta))
\end{equation}
where $\mathcal{W}$ operates independently of $\theta$, so the generative process $v_i(x, \theta)$ remains unaffected by the watermarking mechanism. This isolation of $\mathcal{W}$ from $\theta$ increases the adversarial difficulty, as bypassing or altering $\mathcal{W}$ would necessitate tampering with an external component rather than the model’s core parameters.

\subsubsection{Accessibility}

Accessibility in AIGC watermarking defines the watermark’s security, ensuring that only authorized users can verify it. 
In some AIGC watermarking techniques, access to specific internal components, such as token logits from the language model API and the original prompt, is required to identify the watermark within generated content, to ensure that watermark verification is controlled and secure.

An accessible watermarking scheme requires that the watermark be detectable only if:
\begin{equation}
m' = m \;  \quad   \text{if and only if} \;  \quad  \text{\textit{id}} \in \mathcal{A}
\end{equation}
where $\mathcal{A}$ is the set of authorized \textit{id}s required for watermark detection. 
This restriction makes it challenging for unauthorized users to detect or manipulate the watermark, thus securing the watermark’s integrity.

\section{Functionality of AIGC Watermarking}\label{sec:Functionality}

In this section, we recap the AIGC watermarking techniques from the functionality side and provide the formal definitions of these functionalities. 

\begin{figure}[t]
    \centering
    \includegraphics[width=0.85\linewidth]{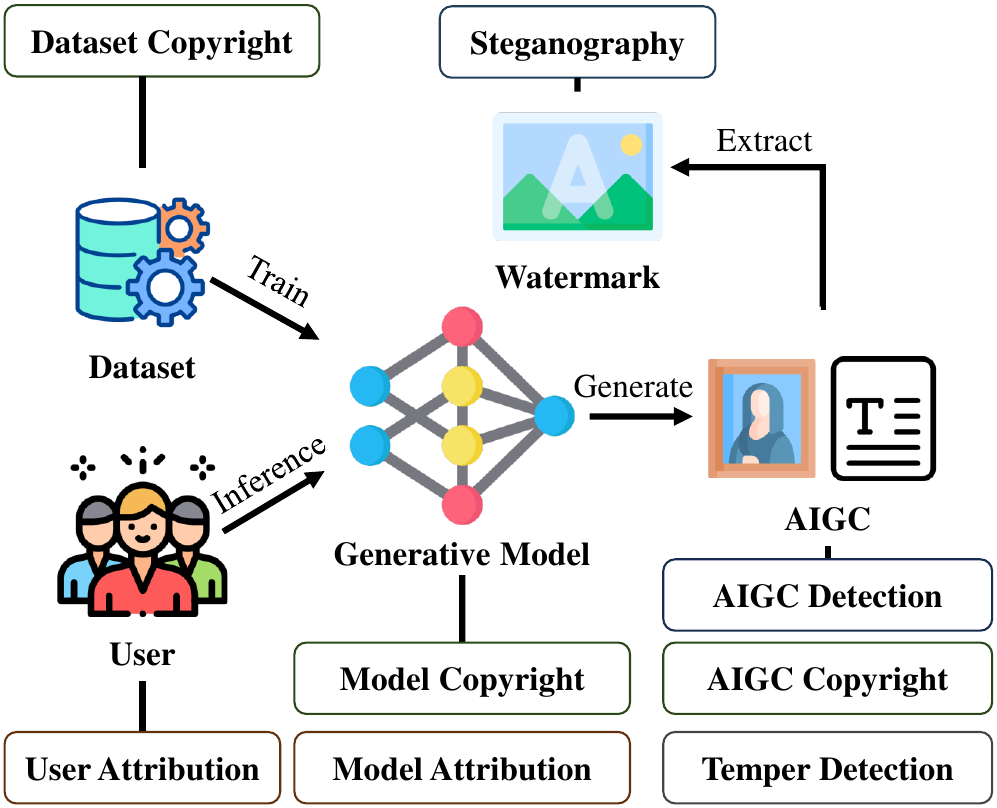}
    \caption{The overview of the functionalities of AIGC watermarking.}
    \label{fig:functionality}
    \vspace{-1em}
\end{figure}

The functionalities of the AIGC watermark can be summarized as five aspects, including detection, attribution, copyright protection, steganography, and tamper detection. We can further refine the functionalities according to the five elements (i.e., dataset, user, generative model, AIGC, and watermark) in the AIGC watermarking. The overview is depicted in \cref{fig:functionality}. Common properties required for AIGC watermarking are stealthiness, soundness, and efficiency.

\subsection{Detection}
One of the key functionalities of watermarking technology is to efficiently and accurately detect AIGC, determining whether a particular information media has been generated by AI. 
AIGC is challenging to distinguish through human judgment \cite{lu2024seeing} with high deceptiveness, leading to various illegal activities, such as AI-enabled fraud and public opinion manipulation. 
This section categorizes methods for enabling AIGC detection in watermarking technology into two types: detection by watermarks embedded during content generation and detection by watermarks embedded directly into the generated content.


\subsubsection{Detection by watermarks embedded during content generation}
This kind of method identifies AIGC by embedding detectable features within the generated outputs during the model's generation phase \cite{zeng2023securing,kirchenbauer2023watermark,liu2019novel,fang2023flow,wang2023free,zhao2023provable,xiong2023flexible}. This approach focuses on embedding watermarks into the nodes other than the root node $r$ in the supply chain, typically by modifying the model parameters \cite{zeng2023securing,kirchenbauer2023watermark}  or altering the training dataset \cite{bui2023rosteals,luo2022leca}. By doing so, the model's outputs inherently carry these embedded markers, which serve as identifiable signatures for detection. Specifically, during the training process, targeted interventions, such as adjusting weights or introducing specialized training samples, encode detectable and uniquely identifiable features into the outputs. These features can then be extracted during the detection phase, allowing the identification of AIGC based on the presence or absence of these embedded watermarks. The core idea is to ensure that any content generated by the model contains detectable markers, making it possible to reliably identify AIGC.

\subsubsection{Detection by watermarks embedded directly to the generated content}
This kind of method often employs a separate watermarking system to process and embed detectable features directly into the AIGC, i.e., the root node $r$ in the supply chain~\cite{singh2017dwt,yang2022tracing,tan2024waterdiff,luo2020distortion,ahmadi2020redmark}. This reduces the reliance of detection methods on the generative model itself, providing greater flexibility and independence. 
Although this approach decouples the detection process from model training, it can still utilize generative AI to enhance the quality of watermarked AIGC and improve detection robustness. For example, replacing traditional CNN-based detection methods with GANs \cite{wen2019romark,liu2017coverless} allows post-processing methods to benefit from enhanced generative capabilities, thereby refining the detection process.



\subsection{Attribution}

The watermark inside the AIGC can also signify its origin. Therefore, AIGC watermarks have the capability named attribution (or traceability) to help trace the origin of the AIGC. Based on the target, attributions can be divided into \emph{model attribution} and \emph{user attribution}. To achieve multi-model or multi-user attribution, AIGC watermarking for attribution needs adequate capability and robustness.

\subsubsection{Model Attribution}

Model attribution aims to identify the generative model responsible for specific content~\cite{liu2024model}. By embedding watermarks within models, the watermark can later be extracted from generated content to verify the model’s origin \cite{zhou2024traceablespeech, lukas2023ptw, fei2022supervised}.

For image models, Lukas et al. \cite{lukas2023ptw} proposed fine-tuning pre-trained generators to produce watermarked images. On the other hand, Fei et al.~\cite{fei2022supervised} insert a pre-trained watermarking decoding block at the output of the GAN generator to achieve model attribution. For text models, Baldassini et al. \cite{baldassini2024cross} utilized cross-attention mechanisms to embed watermarks in large language models (LLMs), introducing methods that minimally impact performance. Qu et al. \cite{qu2024provably} proposed segment-based watermarking with error correction for robust text watermarking. In multi-modal models, Zhou et al. \cite{zhou2024traceablespeech} focused on embedding watermarks to text-to-speech models so that synthetic speech output by the models will contain a watermark. \cite{zhou2024traceablespeech} also further enhanced both traceability and watermark imperceptibility for watermarking multi-modal models.

While the aforementioned studies have provided some initial insights into the issue of model attribution, it is important to acknowledge that a comprehensive solution to model attribution remains unresolved. Furthermore, there is an urgent need for research on different modal models, such as audio, in order to address this gap in the literature.

\subsubsection{User Attribution}

User attribution identifies the individual using AI to generate content. Unlike model attribution, which employs a fixed watermarking module, user attribution requires a flexible system that embeds user-specific messages during generation.

One common approach is fine-tuning models into distinct versions \cite{fernandez2023stable}, where each version generates content with a different watermark for various users. For instance, Fernandez et al. fine-tuned the VAE decoder of latent diffusion models with a pre-trained watermark decoder and fixed watermark message. Kim et al. \cite{kim2024wouaf} modified generative models using the distributor's unique digital fingerprint through weight modulation. However, such approaches can be costly when many users are involved since the overhead of fine-tuning a model is substantial.

To tackle the above limitation, a promising solution involves adding a plug-and-play watermarking module that can embed user-specific watermarks during generation. Feng et al. \cite{feng2023catch} introduced Concept Watermarking to trace malicious users generating illegal images, embedding watermark information into the target concept. GROOT \cite{liu2024groot} encodes watermarks into the latent space for audio synthesis. Overall, flexible watermarking modules can enhance user attribution, enabling more effective accountability in content generation.

\subsection{Copyright Protection}





Copyright protection is a fundamental functionality of digital watermarks. By extracting watermarks from digital assets, the owner can verify their ownership. AIGC watermarking can also be used to protect the copyright of various components in AI, such as datasets, models, and AIGC. 

\subsubsection{Dataset Copyright Protection}

Dataset copyright protection focuses on safeguarding artifacts and datasets from unauthorized use, particularly in training generative models~\cite{li2022untargeted, guo2023domain, li2023black}. Although some cutting-edge works have explored the issue of dataset copyright protection~\cite{wei2024pointncbw, guo2024zero, he2024difficulty}, they are primarily focused on classification models and fail to directly apply to generative models due to the inherent differences in the output formats between generative models and classification models.

As such, AIGC watermarking stands as a promising solution to dataset copyright protection for generative models. It involves two main steps. First, the owner embeds a specific watermark into their artifacts. When these artifacts are used to train a model, the model produces watermarked content. By extracting the watermark from the generated content, the owner can verify whether the model developer has infringed upon their copyright \cite{ye2024duaw, cui2023diffusionshield, zhu2024watermark}. For instance, SIREN~\cite{li2025towards} proposes to optimize the watermark perturbation in a way to be recognized by the model as a feature relevant to the personalization task. For text generation models, WATERFALL \cite{lau2024protecting} is a robust and scalable watermarking framework that operates without requiring training. It integrates a large language model (LLM) paraphraser to embed watermarks in text, ensuring that the watermarks remain verifiable even after paraphrasing, translation, or use in LLM training.

\subsubsection{Model Copyright Protection}

Model copyright protection refers to determining the ownership of a suspicious model~\cite{adi2018turning, shao2025explanation}. Unlike traditional model watermarking methods that conduct ownership verification through testing specific input-output pairs~\cite{li2022defending, yang2023watermarking, shao2024fedtracker}, AIGC watermarking for model copyright protection involves extracting the watermark only from the generated contents~\cite{wu2020watermarking, yang2024gaussian, zhao2023recipe}.

A typical solution to AIGC-watermarking-based model copyright protection is to utilize a specific watermark decoder during training. The watermark decoder can extract the watermark from the AIGC and this watermarking objective is formulated into the training loss function. For instance, Zhang et al. \cite{zhang2020model} proposed embedding invisible watermarks into outputs of image processing models, enabling attackers to learn the watermark when training surrogate models. Wu et al. \cite{wu2020watermarking} developed a joint training approach for host and watermark-extraction networks, achieving a low bit error rate but at a high cost. Peng et al.~\cite{peng2023intellectual} introduced WDM for diffusion models, which encodes watermarks during task generation. Building on the prior works, a few existing works also provide concrete steps to improve the robustness or stealthiness of the AIGC watermarking for model copyright protection~\cite{zhang2024suppressing, liu2023watermarking}.



\subsubsection{AIGC Copyright Protection}

AIGC copyright protection involves legal measures that safeguard the rights of the creators of AIGC. 
Existing AIGC watermarking techniques (e.g., the watermarking methods for detection or attribution)~\cite{kirchenbauer2023watermark, wen2023tree} are instrumental in AIGC copyright protection. By embedding unique identifiers in generated content, watermarking asserts ownership and facilitates tracing of the work's origin. These watermarks can serve as legal evidence, deterring infringement and ensuring the integrity of AIGC against unauthorized modifications. 

\subsection{Steganography}
Steganography focuses on the concealment of secret information within the generated content. In AIGC, steganography enables secure, undetectable embedding of information across multiple modalities, preserving content quality and supporting applications needing confidentiality, authentication, or traceability. The key properties of steganography are capacity and stealthiness.

\subsubsection{Text-based Steganography}

Text-based steganography in AIGC embeds secret information within the generated text, enabling secure data transfer. Methods involve manipulating text features, leveraging natural language patterns, and using advanced LLMs for better capacity and concealment. Early LSTM and RNN approaches \cite{fang2017generating,yang2018rnn} struggled with extraction accuracy and robustness. Liu et al. \cite{liu2024generative} used a ResNet-LSTM to embed information in live comments, achieving high stealth but limited. Recent LLM-based methods like LLM-Stega \cite{wu2024generative} and PPLM-Stega \cite{cao2022generative} also enhance capacity and text quality. 

\subsubsection{Image-based Steganography}

Image-based steganography in AIGC enhances the capacity and security of hidden information within images, often using Generative Adversarial Networks (GANs). Hayes and Danezis \cite{hayes2017generating} embedded secret data into images during generation via adversarial training, achieving robustness against detection tools. Zhang et al. \cite{zhang2019steganogan} introduced \textit{SteganoGAN}, hiding data within images using an encoder, a decoder, and critic networks, attaining high embedding rates up to 4.4 bits per pixel while maintaining low detection rates. Zhang et al. \cite{zhang2023steganography} exploited generative model volatility in black-box scenarios to enhance security. 

\subsubsection{Audio and Video Steganography}

Audio and video steganography embed secret information within auditory and visual media, enabling secure and imperceptible communication. In audio steganography, methods include optimizing resource efficiency, using generative adversarial networks (GANs) for cover audio generation, and coverless approaches that generate steganographic audio directly from secret messages. For instance, Jiang et al. \cite{jiang2020smartsteganogaphy} proposed \textit{SmartSteganography}, a lightweight model for IoT applications that efficiently embeds messages while maintaining audio quality. Chen et al. \cite{chen2021learning} developed a GAN-based framework generating high-quality cover audio resistant to steganalysis, while Li et al. \cite{li2023coverless} introduced a coverless method using WaveGAN to enhance concealment. Video steganography leverages temporal and spatial information in videos to conceal data within semantic attributes of content. Mao et al. \cite{mao2024covert} presented the Robust Generative Video Steganography Network (RoGVSN), encoding secret information into features like facial attributes with high extraction accuracy under distortion attacks. 


\subsection{Tamper Detection}
Tamper detection can be broadly categorized into two aspects: determining whether a watermark-protected sample has been altered and identifying the specific regions within the sample that have undergone modifications. Different from other functionalities, tamper detection depends on the fragility of AIGC watermarking. 

\subsubsection{Sample-Level Tamper Detection}
The primary objective at the sample level is to ascertain the integrity of the entire content by detecting any unauthorized modifications. Common methods for sample-level tamper detection include embedding robust adversarial watermarks, such as Cmua-watermark and Robust Invisible Watermarking (RIW), to ensure that any unauthorized modifications in the sample can be detected effectively. Huang et al.~\cite{huang2022cmua} introduce \textit{Cmua-watermark}, a cross-model universal adversarial watermark designed to disrupt tampered images generated by AI. By embedding adversarial watermarks, their approach ensures that any tampering attempts by deepfake technologies can be effectively identified. Similarly, Tan et al.~\cite{tan2023somewhat} propose \textit{Robust Invisible Watermarking (RIW)}, which leverages adversarial example techniques to embed watermarks that remain extractable even after diffusion-based editing. Their experimental results demonstrate a high extraction accuracy of 96\% post-editing, significantly outperforming traditional methods. These approaches emphasize the importance of embedding robust watermarks that can reliably indicate the integrity of the entire sample.

\subsubsection{Localization of Tampered Regions}
Beyond merely detecting tampering, it is crucial to pinpoint the exact regions within a sample that have been altered. Methods for localizing tampered regions commonly involve embedding dual invisible watermarks. Zhang et al.~\cite{zhang2024editguard} present \textit{EditGuard}, a versatile watermarking framework that embeds dual invisible watermarks to enable precise localization of tampered areas alongside copyright protection. EditGuard utilizes a joint image-bit steganography network (IBSN). Building on this, Zhang et al.~\cite{zhang2024v2a} introduce \textit{V2A-Mark}, which extends watermarking to visual-audio content.

Another solution is to employ robust invertible steganography to preserve watermark integrity, enabling accurate detection of altered areas. Xu et al.~\cite{xu2022robust} develop \textit{Robust Invertible Image Steganography (RIIS)}, which employs a conditional normalizing flow to model high-frequency image components, thereby enhancing the robustness and imperceptibility of hidden watermarks. Such a method collectively advance the capability to not only detect tampering but also accurately identify and localize the modified regions within watermarked content. 


\subsection{Future Work on Functionality of AIGC Watermarking}
\label{sec:fw-watermark}

The field of AIGC watermarking presents several avenues for future research aimed at enhancing security and usability, as follows. 

\subsubsection{Publicly Detectable AIGC Watermarking}
Future research may emphasize the property of public accessibility and the creation of publicly detectable watermarks that enable third-party verification of the origin and authenticity of generated content. Existing works tend to require certain prior knowledge of the embedded watermark (e.g., the watermark decoder or extractor) when extracting the watermark. Developing a publicly detectable AIGC watermarking method aims to ensure that the watermarks can be easily identified by external parties and the public, without necessitating proprietary detection tools. This approach may improve the transparency of AIGC watermarking. 



\subsubsection{Protecting AIGC from Unauthorized Usage for Training}
Protecting AIGC from unauthorized usage for training purposes may also be a crucial area for future exploration while Existing methods mainly focus on the protection of artifacts. Research may focus on developing methods to identify and restrict the use of AIGC in subsequent nodes in the supply chain, such as training datasets. This may help better safeguard the copyright of AIGC and avoid unsafe usage.

\subsubsection{Tamper Detection of Model and Dataset via AIGC Watermarking}
Future work may also address the integration of tamper detection mechanisms for both models and datasets through AIGC watermarking while existing methods only pay attention to AIGC tamper detection. By embedding unique identifiers or verification codes within model architectures or training datasets, researchers could facilitate the tracking of unauthorized modifications or alterations. This enhancement would enhance the overall integrity of AIGC systems, providing a reliable means of verifying the legitimacy of both the models and the data used for training.

\section{Security of AIGC Watermarking}\label{sec:security}

In this section, we first derive the detailed attack surfaces of the AIGC watermark based on the threat model defined in \cref{sec:problem:threat_model}. After that, we introduce the existing attack on the AIGC watermark. Finally, we discuss the potential directions for future research on the security side of the AIGC watermark.

\subsection{Attack Surfaces of AIGC Watermarking}

Recent research \cite{jiang2023evading, zhao2023invisible} indicates that watermarks exhibit vulnerability when faced with attacks. In \cref{sec:problem:threat_model}, we briefly introduced the attack surfaces related to watermarking in AIGC. Here, we delve deeper to gain a better understanding of the relationship between the AIGC watermarking process, the attack surfaces, and the corresponding attack methods.

\begin{itemize}[leftmargin=*]
    \item Manipulating $E$ could result in watermark detection attacks, as it plays a crucial role in the watermark extraction process. Adversaries can search for the watermark's presence, and if it is absent, they can use the AIGC outputs freely, resulting in the misuse of AIGC technologies.
    \item The vulnerability of the watermark inserting function $v'_e$ makes it susceptible to watermark forgery attacks since it is responsible for the watermark embedding process. If adversaries successfully exploit this weakness, they can create counterfeit watermarks that mimic the original. This allows the adversary to falsely attribute AIGC content to the victim, potentially leading to defamation and damage to the victim's reputation.
    \item The manipulation of $x_i$, $v_i$ and $o_i'$ can be grouped together as they occur in the stage where the watermark has been embedded but not yet extracted. These attack surfaces are particularly susceptible to watermark removal attacks. Attackers can adjust model parameters, alter the model structures, or transform outputs to evade the watermark from generated content. Once the watermark is removed, attackers can exploit AIGC content freely, resulting in disseminating false information and infringing on others' intellectual property rights.
\end{itemize}

These attacks raise concerns about the reliability of watermarking as a protective mechanism, emphasizing the importance of developing more robust and resilient watermarking solutions.

\subsection{Watermark Detection}
In watermark detection, attackers aim to detect if the content contains any hidden watermark. Watermark detection attacks specifically target the watermark extractor function $E$, seeking to identify and exploit vulnerabilities in how watermarks are embedded and retrieved. 

As far as we know, there are currently only two works on watermark detection. One approach leverages offset learning to isolate the influence of watermarks by comparing watermarked and non-watermarked datasets, allowing for more precise identification of the embedded watermark signals \cite{pan2024finding}. Another method finds that Enhancing feature extraction for steganographic image detection further improves the ability to detect watermarks, outperforming traditional CNN models \cite{tan2014stacked}. Given the lack of work in this area, it is essential that more research on watermark detection is conducted to advance this field.

\subsection{Watermark Forgery} 
Watermark forgery involves adversaries forging a watermark, making the vanilla content with the forged watermark verified as the watermarked content. These attacks specifically target the watermark embedding function $v'_e$ as the attack surface. In text watermark forgery, techniques often concentrate on analyzing token distributions or utilizing the inherent characteristics of large language models, whereas image watermark forgery generally involves using datasets of source images and watermarked images for adversarial training.

For text watermark forgery, one approach focuses on analyzing token distributions in large language model responses, allowing attackers to approximate watermarking rules and successfully spoof or manipulate watermarks, achieving high success rates \cite{jovanovic2024watermark}. Besides, attackers can exploit the inherent properties of watermarking methods, such as robustness and quality preservation, and manipulate large language models by either removing or spoofing watermarks, further exposing vulnerabilities in the detection systems and compromising the authenticity of AIGC \cite{pang2024attacking}. 

For image watermark forgery, an approach develops a watermark faker using adversarial learning to generate fake watermarked images, but it assumes attackers can access paired original and watermarked images, which is often unrealistic \cite{wang2021watermark}. 

However, current research on watermark forgery remains limited, indicating a need for more comprehensive studies to thoroughly assess the security of watermark against forgery.

\subsection{Watermark Removal}
In watermark removal, attackers aim to erase the watermark hidden in the AIGC while preserving its visual quality. These attacks specifically target the watermarked content, meaning that $x_i$, $v_i$, and $o'_i$ could potentially serve as attack surfaces. However, due to the limited capabilities, attackers typically only have access to the final output $y'$. As a result, existing attacks typically view the final output $y'$ as the sole attack surface. Current watermark removal attacks can be categorized into several types:

\begin{itemize}[leftmargin=*]
    \item Regeneration attacks \cite{zhao2023generative, zhao2023invisible, he2024can}: These methods focus on regenerating images or text to eliminate watermarks.
    \item Adversarial attacks \cite{jiang2023evading}: These techniques manipulate the content using adversarial perturbations to evade detection.
    \item Green token minimization attacks \cite{wu2024bypassing}: This category specifically targets text watermarks, where attackers aim to minimize the presence of green tokens that signify the presence of a watermark.
    \item Editing attacks: This category includes various editing techniques, for example, JPEG compression for images.
\end{itemize}

For image watermark removal, utilizing an adversarial optimization strategy can evade detection in both white-box and black-box scenarios effectively \cite{jiang2023evading}. Random walk achieves successful watermark removal by reducing detection accuracy through iterative modifications \cite{zhang2023watermarks}. Adversarial methods targeting spectral traces within watermarked images can also effectively disrupt various watermarking schemes, though this may result in slight degradation of visual quality \cite{kassis2024unmarker}. Moreover, Zhao et al.~\cite{zhao2023generative, zhao2023invisible} present a regeneration attack that targets invisible watermarks by degrading watermarked images into embeddings, adding random noise to disrupt the watermark, and then reconstructing the image from the altered embeddings.

In the realm of text watermark removal, He et al. propose the Cross-lingual Watermark Removal Attack (CWRA), which translates watermarked text to a pivot language and back, reducing watermark strength, with a limitation on the dependence of high-quality machine translation \cite{he2024can}. 
In addition, Wu et al. propose a method to evade watermark detection in LLM-generated text by prompting the LLM to generate text fragments that resemble human writing while minimizing green tokens, thus neutralizing the watermark \cite{wu2024bypassing}. Furthermore, Chang et al. present a watermark removal attack that uses a weaker reference LLM to infer watermarked tokens and smooth distribution perturbations. However, this approach relies on multiple queries to the watermarked model, which may be costly or impractical in some scenarios \cite{chang2024watermark}. Zhang et al. propose a method to remove watermarks from LLM-generated text by leveraging watermark detection rules and modeling green list theft as a mixed integer programming problem \cite{zhang2024large}.

Among all those attack types, we believe regeneration attacks show significant promise due to their less stringent assumptions about attackers' capabilities and their demonstrated effectiveness in removing watermarks with minimal perturbations to the content. However, current regeneration attacks perform poorly against certain watermark methods \cite{fernandez2022watermarking, tancik2020stegastamp}, which highlights a critical area for future research and improvement.

\subsection{Future Work on AIGC Watermark Security}
Despite the growing interest in watermarking technologies, research targeting the security of watermark remains limited. Thus, we believe the field is still in its nascent stages. In this section, we propose several potential research directions that could inspire the future development of watermark attacks, paving the way for more robust and secure watermarking solutions.

\subsubsection{More Complex Defense for Endogenous Watermarks}
Different from traditional watermarks, endogeneity has become mainstream to embed watermarks into AIGC. However, this may introduce more complex defense update mechanisms. Instead of directly modifying the small watermark module, the defense update for endogenous watermarks probably modifies the whole generative model whose overhead is unacceptable, especially for large models. Hence, it is essential to study more low-cost defense mechanisms for endogenous watermarks.

\subsubsection{Security on Cross-Modality}
Most existing watermark attacks and defenses target images or text.
However, as the progress of cross-modality in generative models, especially multi-modality interaction in large models, their security remains unknown. Advanced LLMs like GPT-4o introduce HER mode, which allows users to employ both text and voice to interact with the model. Underlying the interactions, the tokens for text and voice are fused to produce outputs, introducing new attack surfaces to the corresponding watermarking modules that need further study.

\subsubsection{Attack against Semantic Watermarking}
Most existing watermark attacks \cite{zhao2023generative, zhao2023invisible} fail when confronted with semantic watermarking \cite{wen2023tree}. This gap indicates a critical area for future study, as the robustness of semantic watermarks remains largely untested. Addressing the vulnerabilities of semantic watermarking should be a priority for researchers, as understanding and developing attack methods could lead to improved watermarking techniques and enhanced security measures.

\section{AIGC Governance}\label{sec:Regulations}
The widespread adoption of AIGC services raises concerns regarding the safety and trustworthiness of AI-generated content and draws significant attention to the governance of such services from countries and organizations.
In this section, we first introduce the AIGC regulations from the aspect of various countries, followed by a more detailed examination of AIGC regulation from the perspective of five key elements of AIGC watermarking, dataset, user, generative model, AI-generated content, and watermark. 
Finally, we will discuss the future of AIGC governance.

\subsection{AIGC Governance Status}
Recently, countries and organizations around the world have begun to formulate laws and regulations to govern AIGC and related services. The United States has released Blueprint for an AI Bill of Rights \cite{blueprint}, which emphasizes issues of fairness, transparency, privacy protection, and the development of AI. Besides, The AI Risk Management Framework \cite{riskframework} aims to regulate the use of AI, while the Malicious Deep Fake Prohibition Act \cite{MaliciousAct} seeks to prohibit the malicious use of deepfake technology. Despite the regulations above, the United States has not yet proposed a legal framework specifically for AIGC. 
The European Union has passed the EU Artificial Intelligence Act \cite{euact}, which addresses the safety, ethics, and fairness of AI, emphasizing the impact of AI technology on privacy rights and personal data, and focusing on the protection of users and consumers. China places significant emphasis on the control of AIGC, having released a series of regulations ~\cite{gccontentregulation,draft,GBdraft}, a practice guideline~\cite{CybersecurityStandards}, and a compulsory standard \cite{GB} for AIGC watermarking. China focuses more on the impact of AIGC on social governance and public safety, with stronger scrutiny and restrictions regarding the ethical and moral aspects of AIGC.

\subsection{AIGC Regulations From Five Elements}

\subsubsection{Dataset} \leavevmode

\noindent \emph{\textbf{Data processing:}} As the dataset to train the AI model can be accessed publicly or privately, suffering from the risk of potential privacy leaks, the \cite{GDPR} requires that privacy should be protected by data anonymization and data encryption during data processing. Besides, low-quality data that is ambitious, violating, biased, or politically sensitive needs to be filtered before being used.
\cite{Directive} also requires the collection of datasets to obey the regulations on Intellectual protection and business competition.

\noindent \emph{\textbf{Cross-border data flow:}} Singapore and China have restrictions \cite{PDPA,PIPL} on the storage of related data to be stored entirely within the country's borders.

\subsubsection{Generative model} \leavevmode

\noindent \emph{\textbf{Open source:}} Whether AI models should be open-sourced, and to what extent, has long been a controversial issue. Several regulations have called for AI model structures and algorithms to be transparent to the public to foster trust and accountability. While open-sourcing models can drive technological innovation and ensure transparency, they also raise concerns about the potential misuse of these models by malicious individuals \cite{doi:10.1126/science.adp1848}.

\subsubsection{AI-generated content} \leavevmode

\noindent \emph{\textbf{Sensitive content filtering:}}sensitive content is required to be filtered from both user prompts and model outputs to make sure that the AIGC complies with local regulations and to prevent AIGC from being used for illegal purposes \cite{Digitalservicesact}.

\subsubsection{Watermark} \leavevmode

\noindent \emph{\textbf{Explicit indication:}}
 In scenarios, such as voice synthesis and face generation, an explicit marker is required to be prominently displayed at appropriate locations or regions within the generated content to inform the public of the deep synthesis involved \cite{gccontentregulation,draft}.

\noindent \emph{\textbf{AIGC tracing:}}
Watermarks need to be capable of tracing the identity of users in cases where AIGC is misused \cite{euact,GB}.
Some countries also require that user identity tracing methods must be robust enough to withstand potential attacks.

\subsubsection{User} \leavevmode

\noindent \emph{\textbf{Content compliance:}} User should not utilize AIGC that does not comply with local laws or include pornographic, violent, or bloody content.
Some countries especially stress the legal applications in terms of politics or religion \cite{euact,Ecological}. 
Spreading AI-generated false or misleading content also carries legal risks.

\noindent \emph{\textbf{Intellectual protection:}}
The intellectual property rights of AIGC remain controversial, where regulations in different countries hold different interpretations.
It is still hard to determine the ownership of AIGC copyrights, which should belong to the source data owner, the model developer, the AI itself, or the AIGC service users \cite{euact}.

\subsection{Future Work on AIGC Governance}
The regulations and laws currently issued by various countries have begun to outline the future landscape of AIGC governance. However, the formulation of AIGC regulation still faces numerous challenges. For instance, issues related to intellectual property—specifically, how to define whether AI-generated content belongs to the user or the AI developer—remain unresolved. Additionally, the question of liability—how responsibility should be attributed when AIGC leads to issues such as infringement or defamation—has yet to be clearly defined. Furthermore, due to differences in regulatory approaches, values, technical understanding, and legal systems across countries, there is a lack of a unified global regulatory framework for AIGC, which hinders the innovation and development of AI techniques.

\section{Conclusion}

Watermark flows in the supply chain (of AIGC).
We call for zooming out the scope of watermarking from the content itself to the supply chain, and thus see its impact on the entire system.
We provide a formal definition of the AIGC watermark to distinguish it from the traditional watermark.
We span the discussion of functionality and security of AIGC watermark from the perspective of AIGC watermark properties.
We encourage the community to comprehensively explore the properties of the AIGC watermark and find the potential research directions. 
We hope that this work serves as a summary of current AIGC watermarks and boosts the research in the field of AIGC watermarks.


\bibliographystyle{IEEEtran}
{\small \bibliography{IEEEabrv, paper}}

\begin{thebibliography}{100}
\providecommand{\url}[1]{#1}
\csname url@samestyle\endcsname
\providecommand{\newblock}{\relax}
\providecommand{\bibinfo}[2]{#2}
\providecommand{\BIBentrySTDinterwordspacing}{\spaceskip=0pt\relax}
\providecommand{\BIBentryALTinterwordstretchfactor}{4}
\providecommand{\BIBentryALTinterwordspacing}{\spaceskip=\fontdimen2\font plus
\BIBentryALTinterwordstretchfactor\fontdimen3\font minus \fontdimen4\font\relax}
\providecommand{\BIBforeignlanguage}[2]{{%
\expandafter\ifx\csname l@#1\endcsname\relax
\typeout{** WARNING: IEEEtran.bst: No hyphenation pattern has been}%
\typeout{** loaded for the language `#1'. Using the pattern for}%
\typeout{** the default language instead.}%
\else
\language=\csname l@#1\endcsname
\fi
#2}}
\providecommand{\BIBdecl}{\relax}
\BIBdecl

\bibitem{achiam2023gpt}
J.~Achiam, S.~Adler, S.~Agarwal, L.~Ahmad, I.~Akkaya, F.~L. Aleman, D.~Almeida, J.~Altenschmidt, S.~Altman, S.~Anadkat \emph{et~al.}, ``Gpt-4 technical report,'' \emph{arXiv preprint arXiv:2303.08774}, 2023.

\bibitem{rombach2022high}
R.~Rombach, A.~Blattmann, D.~Lorenz, P.~Esser, and B.~Ommer, ``High-resolution image synthesis with latent diffusion models,'' in \emph{Proceedings of the IEEE/CVF conference on computer vision and pattern recognition}, 2022, pp. 10\,684--10\,695.

\bibitem{touvron2023llama}
H.~Touvron, T.~Lavril, G.~Izacard, X.~Martinet, M.-A. Lachaux, T.~Lacroix, B.~Rozi{\`e}re, N.~Goyal, E.~Hambro, F.~Azhar \emph{et~al.}, ``Llama: Open and efficient foundation language models,'' \emph{arXiv preprint arXiv:2302.13971}, 2023.

\bibitem{thirunavukarasu2023large}
A.~J. Thirunavukarasu, D.~S.~J. Ting, K.~Elangovan, L.~Gutierrez, T.~F. Tan, and D.~S.~W. Ting, ``Large language models in medicine,'' \emph{Nature medicine}, vol.~29, no.~8, pp. 1930--1940, 2023.

\bibitem{xiao2024omnigen}
S.~Xiao, Y.~Wang, J.~Zhou, H.~Yuan, X.~Xing, R.~Yan, S.~Wang, T.~Huang, and Z.~Liu, ``Omnigen: Unified image generation,'' \emph{arXiv preprint arXiv:2409.11340}, 2024.

\bibitem{pang2024attacking}
Q.~Pang, S.~Hu, W.~Zheng, and V.~Smith, ``Attacking llm watermarks by exploiting their strengths,'' \emph{arXiv preprint arXiv:2402.16187}, 2024.

\bibitem{jovanovic2024watermark}
N.~Jovanovi{\'c}, R.~Staab, and M.~Vechev, ``Watermark stealing in large language models,'' \emph{arXiv preprint arXiv:2402.19361}, 2024.

\bibitem{zhang2023watermarks}
H.~Zhang, B.~L. Edelman, D.~Francati, D.~Venturi, G.~Ateniese, and B.~Barak, ``Watermarks in the sand: Impossibility of strong watermarking for generative models,'' \emph{arXiv preprint arXiv:2311.04378}, 2023.

\bibitem{jiang2023evading}
Z.~Jiang, J.~Zhang, and N.~Z. Gong, ``Evading watermark based detection of ai-generated content,'' in \emph{Proceedings of the 2023 ACM SIGSAC Conference on Computer and Communications Security}, 2023, pp. 1168--1181.

\bibitem{zhang2024v2a}
X.~Zhang, Y.~Xu, R.~Li, J.~Yu, W.~Li, Z.~Xu, and J.~Zhang, ``V2a-mark: Versatile deep visual-audio watermarking for manipulation localization and copyright protection,'' \emph{arXiv preprint arXiv:2404.16824}, 2024.

\bibitem{zhang2024editguard}
X.~Zhang, R.~Li, J.~Yu, Y.~Xu, W.~Li, and J.~Zhang, ``Editguard: Versatile image watermarking for tamper localization and copyright protection,'' in \emph{Proceedings of the IEEE/CVF Conference on Computer Vision and Pattern Recognition}, 2024, pp. 11\,964--11\,974.

\bibitem{liu2024survey}
A.~Liu, L.~Pan, Y.~Lu, J.~Li, X.~Hu, X.~Zhang, L.~Wen, I.~King, H.~Xiong, and P.~Yu, ``A survey of text watermarking in the era of large language models,'' \emph{ACM Computing Surveys}, vol.~57, no.~2, pp. 1--36, 2024.

\bibitem{hosny2024digital}
K.~M. Hosny, A.~Magdi, O.~ElKomy, and H.~M. Hamza, ``Digital image watermarking using deep learning: A survey,'' \emph{Computer Science Review}, vol.~53, p. 100662, 2024.

\bibitem{zhong2023brief}
X.~Zhong, A.~Das, F.~Alrasheedi, and A.~Tanvir, ``A brief, in-depth survey of deep learning-based image watermarking,'' \emph{Applied Sciences}, vol.~13, no.~21, p. 11852, 2023.

\bibitem{2024Supply}
\BIBentryALTinterwordspacing
``Supply chain attack.'' [Online]. Available: \url{https://en.wikipedia.org/w/index.php?title=Supply_chain_attack&oldid=1246528305}
\BIBentrySTDinterwordspacing

\bibitem{format1}
S.~G. Rizzo, F.~Bertini, and D.~Montesi, ``Content-preserving text watermarking through unicode homoglyph substitution,'' in \emph{Proceedings of the 20th International Database Engineering \& Applications Symposium}, 2016, pp. 97--104.

\bibitem{format2}
J.~T. Brassil, S.~Low, N.~F. Maxemchuk, and L.~O'Gorman, ``Electronic marking and identification techniques to discourage document copying,'' \emph{IEEE Journal on Selected Areas in Communications}, vol.~13, no.~8, pp. 1495--1504, 1995.

\bibitem{format3}
L.~Y. Por, K.~Wong, and K.~O. Chee, ``Unispach: A text-based data hiding method using unicode space characters,'' \emph{Journal of Systems and Software}, vol.~85, no.~5, pp. 1075--1082, 2012.

\bibitem{yang2022tracing}
X.~Yang, J.~Zhang, K.~Chen, W.~Zhang, Z.~Ma, F.~Wang, and N.~Yu, ``Tracing text provenance via context-aware lexical substitution,'' in \emph{Proceedings of the AAAI Conference on Artificial Intelligence}, vol.~36, no.~10, 2022, pp. 11\,613--11\,621.

\bibitem{lexcialsubstitution2}
X.~Yang, K.~Chen, W.~Zhang, C.~Liu, Y.~Qi, J.~Zhang, H.~Fang, and N.~Yu, ``Watermarking text generated by black-box language models,'' \emph{arXiv preprint arXiv:2305.08883}, 2023.

\bibitem{lexcialsubstitution3}
K.~Yoo, W.~Ahn, J.~Jang, and N.~Kwak, ``Robust multi-bit natural language watermarking through invariant features,'' \emph{arXiv preprint arXiv:2305.01904}, 2023.

\bibitem{sir}
\BIBentryALTinterwordspacing
A.~Liu, L.~Pan, X.~Hu, S.~Meng, and L.~Wen, ``A semantic invariant robust watermark for large language models,'' 2024. [Online]. Available: \url{https://arxiv.org/abs/2310.06356}
\BIBentrySTDinterwordspacing

\bibitem{kirchenbauer2023watermark}
J.~Kirchenbauer, J.~Geiping, Y.~Wen, J.~Katz, I.~Miers, and T.~Goldstein, ``A watermark for large language models,'' in \emph{International Conference on Machine Learning}.\hskip 1em plus 0.5em minus 0.4em\relax PMLR, 2023, pp. 17\,061--17\,084.

\bibitem{unforgeable}
A.~Liu, L.~Pan, X.~Hu, S.~Li, L.~Wen, I.~King, and S.~Y. Philip, ``An unforgeable publicly verifiable watermark for large language models,'' in \emph{The Twelfth International Conference on Learning Representations}, 2023.

\bibitem{zhao2023provable}
X.~Zhao, P.~Ananth, L.~Li, and Y.-X. Wang, ``Provable robust watermarking for ai-generated text,'' \emph{arXiv preprint arXiv:2306.17439}, 2023.

\bibitem{kth}
R.~Kuditipudi, J.~Thickstun, T.~Hashimoto, and P.~Liang, ``Robust distortion-free watermarks for language models,'' \emph{arXiv preprint arXiv:2307.15593}, 2023.

\bibitem{aar}
S.~Aaronson and H.~Kirchner., ``Watermarking gpt outputs.'' 2022, https://www.scottaaronson.com/talks/watermark.ppt.

\bibitem{learnability}
C.~Gu, X.~L. Li, P.~Liang, and T.~Hashimoto, ``On the learnability of watermarks for language models,'' \emph{arXiv preprint arXiv:2312.04469}, 2023.

\bibitem{reinforcement}
X.~Xu, Y.~Yao, and Y.~Liu, ``Learning to watermark llm-generated text via reinforcement learning,'' \emph{arXiv preprint arXiv:2403.10553}, 2024.

\bibitem{kalker1999video}
T.~Kalker, G.~Depovere, J.~Haitsma, and M.~J. Maes, ``Video watermarking system for broadcast monitoring,'' in \emph{Security and Watermarking of Multimedia contents}, vol. 3657.\hskip 1em plus 0.5em minus 0.4em\relax SPIE, 1999, pp. 103--112.

\bibitem{chopra2012lsb}
D.~Chopra, P.~Gupta, G.~Sanjay, and A.~Gupta, ``Lsb based digital image watermarking for gray scale image,'' \emph{IOSR journal of Computer Engineering}, vol.~6, no.~1, pp. 36--41, 2012.

\bibitem{ernawan2021improved}
F.~Ernawan, D.~Ariatmanto, and A.~Firdaus, ``An improved image watermarking by modifying selected dwt-dct coefficients,'' \emph{IEEE Access}, vol.~9, pp. 45\,474--45\,485, 2021.

\bibitem{mohammed2023blind}
A.~O. Mohammed, H.~I. Hussein, R.~J. Mstafa, and A.~M. Abdulazeez, ``A blind and robust color image watermarking scheme based on dct and dwt domains,'' \emph{Multimedia Tools and Applications}, vol.~82, no.~21, pp. 32\,855--32\,881, 2023.

\bibitem{wen2023tree}
Y.~Wen, J.~Kirchenbauer, J.~Geiping, and T.~Goldstein, ``Tree-ring watermarks: Fingerprints for diffusion images that are invisible and robust,'' \emph{arXiv preprint arXiv:2305.20030}, 2023.

\bibitem{yang2024gaussian}
Z.~Yang, K.~Zeng, K.~Chen, H.~Fang, W.~Zhang, and N.~Yu, ``Gaussian shading: Provable performance-lossless image watermarking for diffusion models,'' in \emph{Proceedings of the IEEE/CVF Conference on Computer Vision and Pattern Recognition}, 2024, pp. 12\,162--12\,171.

\bibitem{tan2024waterdiff}
Y.~Tan, Y.~Peng, H.~Fang, B.~Chen, and S.-T. Xia, ``Waterdiff: Perceptual image watermarks via diffusion model,'' in \emph{ICASSP 2024-2024 IEEE International Conference on Acoustics, Speech and Signal Processing (ICASSP)}.\hskip 1em plus 0.5em minus 0.4em\relax IEEE, 2024, pp. 3250--3254.

\bibitem{lukas2023ptw}
N.~Lukas and F.~Kerschbaum, ``$\{$PTW$\}$: Pivotal tuning watermarking for $\{$Pre-Trained$\}$ image generators,'' in \emph{USENIX Security Symposium}, 2023, pp. 2241--2258.

\bibitem{zhao2023recipe}
Y.~Zhao, T.~Pang, C.~Du, X.~Yang, N.-M. Cheung, and M.~Lin, ``A recipe for watermarking diffusion models,'' \emph{arXiv preprint arXiv:2303.10137}, 2023.

\bibitem{zeng2023securing}
Y.~Zeng, M.~Zhou, Y.~Xue, and V.~M. Patel, ``Securing deep generative models with universal adversarial signature,'' \emph{arXiv preprint arXiv:2305.16310}, 2023.

\bibitem{fernandez2023stable}
P.~Fernandez, G.~Couairon, H.~J{\'e}gou, M.~Douze, and T.~Furon, ``The stable signature: Rooting watermarks in latent diffusion models,'' in \emph{Proceedings of the IEEE/CVF International Conference on Computer Vision}, 2023, pp. 22\,466--22\,477.

\bibitem{chen2023wavmark}
G.~Chen, Y.~Wu, S.~Liu, T.~Liu, X.~Du, and F.~Wei, ``Wavmark: Watermarking for audio generation,'' \emph{arXiv preprint arXiv:2308.12770}, 2023.

\bibitem{liu2023dear}
C.~Liu, J.~Zhang, H.~Fang, Z.~Ma, W.~Zhang, and N.~Yu, ``Dear: A deep-learning-based audio re-recording resilient watermarking,'' in \emph{Proceedings of the AAAI Conference on Artificial Intelligence}, vol.~37, no.~11, 2023, pp. 13\,201--13\,209.

\bibitem{cheng2024hifi}
X.~Cheng, Y.~Wang, C.~Liu, D.~Hu, and Z.~Su, ``Hifi-ganw: Watermarked speech synthesis via fine-tuning of hifi-gan,'' \emph{IEEE Signal Processing Letters}, 2024.

\bibitem{zhou2024traceablespeech}
J.~Zhou, J.~Yi, T.~Wang, J.~Tao, Y.~Bai, C.~Y. Zhang, Y.~Ren, and Z.~Wen, ``Traceablespeech: Towards proactively traceable text-to-speech with watermarking,'' \emph{arXiv preprint arXiv:2406.04840}, 2024.

\bibitem{san2024latent}
R.~San~Roman, P.~Fernandez, A.~Deleforge, Y.~Adi, and R.~Serizel, ``Latent watermarking of audio generative models,'' 2024.

\bibitem{asikuzzaman2016robust}
M.~Asikuzzaman, M.~J. Alam, A.~J. Lambert, and M.~R. Pickering, ``Robust dt cwt-based dibr 3d video watermarking using chrominance embedding,'' \emph{IEEE Transactions on Multimedia}, vol.~18, no.~9, pp. 1733--1748, 2016.

\bibitem{campisi2005video}
P.~Campisi and A.~Neri, ``Video watermarking in the 3d-dwt domain using perceptual masking,'' in \emph{IEEE international conference on image processing 2005}, vol.~1.\hskip 1em plus 0.5em minus 0.4em\relax IEEE, 2005, pp. I--997.

\bibitem{weng2019high}
X.~Weng, Y.~Li, L.~Chi, and Y.~Mu, ``High-capacity convolutional video steganography with temporal residual modeling,'' in \emph{Proceedings of the 2019 on international conference on multimedia retrieval}, 2019, pp. 87--95.

\bibitem{mishra2019vstegnet}
A.~Mishra, S.~Kumar, A.~Nigam, and S.~Islam, ``Vstegnet: Video steganography network using spatio-temporal features and micro-bottleneck.'' in \emph{BMVC}, vol. 274, 2019.

\bibitem{luo2104dvmark}
X.~Luo, Y.~Li, H.~Chang, C.~Liu, P.~Milanfar, and F.~Yang, ``Dvmark: A deep multiscale framework for video watermarking. arxiv 2021,'' \emph{arXiv preprint arXiv:2104.12734}.

\bibitem{zhang2023novel}
Y.~Zhang, J.~Ni, W.~Su, and X.~Liao, ``A novel deep video watermarking framework with enhanced robustness to h. 264/avc compression,'' in \emph{Proceedings of the 31st ACM International Conference on Multimedia}, 2023, pp. 8095--8104.

\bibitem{mou2023large}
C.~Mou, Y.~Xu, J.~Song, C.~Zhao, B.~Ghanem, and J.~Zhang, ``Large-capacity and flexible video steganography via invertible neural network,'' in \emph{Proceedings of the IEEE/CVF Conference on Computer Vision and Pattern Recognition}, 2023, pp. 22\,606--22\,615.

\bibitem{asikuzzaman2017overview}
M.~Asikuzzaman and M.~R. Pickering, ``An overview of digital video watermarking,'' \emph{IEEE Transactions on Circuits and Systems for Video Technology}, 2017.

\bibitem{synthid}
\BIBentryALTinterwordspacing
DeemMind, ``Synthid: Identifying ai-generated content with synthids,'' 2024. [Online]. Available: \url{https://deepmind.google/technologies/synthid/}
\BIBentrySTDinterwordspacing

\bibitem{guo2024avsecure}
B.~Guo, H.~Tai, G.~Luo, and Y.~Zhu, ``Avsecure: An audio-visual watermarking framework for proactive deepfake detection,'' in \emph{2024 IEEE 14th International Conference on Electronics Information and Emergency Communication (ICEIEC)}.\hskip 1em plus 0.5em minus 0.4em\relax IEEE, 2024, pp. 1--4.

\bibitem{liu2023t2iw}
A.-A. Liu, G.~Zhang, Y.~Su, N.~Xu, Y.~Zhang, and L.~Wang, ``T2iw: Joint text to image \& watermark generation,'' \emph{arXiv preprint arXiv:2309.03815}, 2023.

\bibitem{aberna2024digital}
P.~Aberna and L.~Agilandeeswari, ``Digital image and video watermarking: methodologies, attacks, applications, and future directions,'' \emph{Multimedia Tools and Applications}, 2024.

\bibitem{begum2020digital}
M.~Begum and M.~S. Uddin, ``Digital image watermarking techniques: a review,'' \emph{Information}, 2020.

\bibitem{agarwal2019survey}
N.~Agarwal, A.~K. Singh, and P.~K. Singh, ``Survey of robust and imperceptible watermarking,'' \emph{Multimedia Tools and Applications}, 2019.

\bibitem{wang2023data}
Z.~Wang, O.~Byrnes, H.~Wang, R.~Sun, C.~Ma, H.~Chen, Q.~Wu, and M.~Xue, ``Data hiding with deep learning: A survey unifying digital watermarking and steganography,'' \emph{IEEE Transactions on Computational Social Systems}, 2023.

\bibitem{fairoze2023publicly}
J.~Fairoze, S.~Garg, S.~Jha, S.~Mahloujifar, M.~Mahmoody, and M.~Wang, ``Publicly detectable watermarking for language models,'' \emph{arXiv preprint arXiv:2310.18491}, 2023.

\bibitem{christ2024undetectable}
M.~Christ, S.~Gunn, and O.~Zamir, ``Undetectable watermarks for language models,'' in \emph{The Thirty Seventh Annual Conference on Learning Theory}.\hskip 1em plus 0.5em minus 0.4em\relax PMLR, 2024.

\bibitem{langelaar2000watermarking}
G.~C. Langelaar, I.~Setyawan, and R.~L. Lagendijk, ``Watermarking digital image and video data. a state-of-the-art overview,'' \emph{IEEE Signal processing magazine}, 2000.

\bibitem{anand2021watermarking}
A.~Anand and A.~K. Singh, ``Watermarking techniques for medical data authentication: a survey,'' \emph{Multimedia Tools and Applications}, 2021.

\bibitem{jiao2019review}
S.~Jiao, C.~Zhou, Y.~Shi, W.~Zou, and X.~Li, ``Review on optical image hiding and watermarking techniques,'' \emph{Optics \& Laser Technology}, 2019.

\bibitem{wan2022comprehensive}
W.~Wan, J.~Wang, Y.~Zhang, J.~Li, H.~Yu, and J.~Sun, ``A comprehensive survey on robust image watermarking,'' \emph{Neurocomputing}, 2022.

\bibitem{ma2023generative}
Y.~Ma, Z.~Zhao, X.~He, Z.~Li, M.~Backes, and Y.~Zhang, ``Generative watermarking against unauthorized subject-driven image synthesis,'' \emph{arXiv preprint arXiv:2306.07754}, 2023.

\bibitem{singh2013survey}
P.~Singh and R.~S. Chadha, ``A survey of digital watermarking techniques, applications and attacks,'' \emph{International Journal of Engineering and Innovative Technology (IJEIT)}, 2013.

\bibitem{mohanarathinam2020digital}
A.~Mohanarathinam, S.~Kamalraj, G.~Prasanna~Venkatesan, R.~V. Ravi, and C.~Manikandababu, ``Digital watermarking techniques for image security: a review,'' \emph{Journal of Ambient Intelligence and Humanized Computing}, 2020.

\bibitem{lu2024seeing}
Z.~Lu, D.~Huang, L.~Bai, J.~Qu, C.~Wu, X.~Liu, and W.~Ouyang, ``Seeing is not always believing: benchmarking human and model perception of ai-generated images,'' \emph{Advances in Neural Information Processing Systems}, vol.~36, 2024.

\bibitem{liu2019novel}
Y.~Liu, M.~Guo, J.~Zhang, Y.~Zhu, and X.~Xie, ``A novel two-stage separable deep learning framework for practical blind watermarking,'' in \emph{Proceedings of the 27th ACM International conference on multimedia}, 2019, pp. 1509--1517.

\bibitem{fang2023flow}
H.~Fang, Y.~Qiu, K.~Chen, J.~Zhang, W.~Zhang, and E.-C. Chang, ``Flow-based robust watermarking with invertible noise layer for black-box distortions,'' in \emph{Proceedings of the AAAI conference on artificial intelligence}, vol.~37, no.~4, 2023, pp. 5054--5061.

\bibitem{wang2023free}
R.~Wang, J.~Ren, B.~Li, T.~She, W.~Zhang, L.~Fang, J.~Chen, and L.~Wang, ``Free fine-tuning: A plug-and-play watermarking scheme for deep neural networks,'' in \emph{Proceedings of the 31st ACM International Conference on Multimedia}, 2023, pp. 8463--8474.

\bibitem{xiong2023flexible}
C.~Xiong, C.~Qin, G.~Feng, and X.~Zhang, ``Flexible and secure watermarking for latent diffusion model,'' in \emph{Proceedings of the 31st ACM International Conference on Multimedia}, 2023, pp. 1668--1676.

\bibitem{bui2023rosteals}
T.~Bui, S.~Agarwal, N.~Yu, and J.~Collomosse, ``Rosteals: Robust steganography using autoencoder latent space,'' in \emph{Proceedings of the IEEE/CVF Conference on Computer Vision and Pattern Recognition}, 2023, pp. 933--942.

\bibitem{luo2022leca}
X.~Luo, M.~Goebel, E.~Barshan, and F.~Yang, ``Leca: A learned approach for efficient cover-agnostic watermarking,'' \emph{arXiv preprint arXiv:2206.10813}, 2022.

\bibitem{singh2017dwt}
D.~Singh and S.~K. Singh, ``Dwt-svd and dct based robust and blind watermarking scheme for copyright protection,'' \emph{Multimedia Tools and Applications}, vol.~76, no.~11, pp. 13\,001--13\,024, 2017.

\bibitem{luo2020distortion}
X.~Luo, R.~Zhan, H.~Chang, F.~Yang, and P.~Milanfar, ``Distortion agnostic deep watermarking,'' in \emph{Proceedings of the IEEE/CVF conference on computer vision and pattern recognition}, 2020, pp. 13\,548--13\,557.

\bibitem{ahmadi2020redmark}
M.~Ahmadi, A.~Norouzi, N.~Karimi, S.~Samavi, and A.~Emami, ``Redmark: Framework for residual diffusion watermarking based on deep networks,'' \emph{Expert Systems with Applications}, vol. 146, p. 113157, 2020.

\bibitem{wen2019romark}
B.~Wen and S.~Aydore, ``Romark: A robust watermarking system using adversarial training,'' \emph{arXiv preprint arXiv:1910.01221}, 2019.

\bibitem{liu2017coverless}
M.-m. Liu, M.-q. Zhang, J.~Liu, Y.-n. Zhang, and Y.~Ke, ``Coverless information hiding based on generative adversarial networks,'' \emph{arXiv preprint arXiv:1712.06951}, 2017.

\bibitem{liu2024model}
F.~Liu, H.~Luo, Y.~Li, P.~Torr, and J.~Gu, ``Which model generated this image? a model-agnostic approach for origin attribution,'' in \emph{European Conference on Computer Vision}, 2024.

\bibitem{fei2022supervised}
J.~Fei, Z.~Xia, B.~Tondi, and M.~Barni, ``Supervised gan watermarking for intellectual property protection,'' in \emph{IEEE International Workshop on Information Forensics and Security}.\hskip 1em plus 0.5em minus 0.4em\relax IEEE, 2022, pp. 1--6.

\bibitem{baldassini2024cross}
F.~B. Baldassini, H.~H. Nguyen, C.-C. Chang, and I.~Echizen, ``Cross-attention watermarking of large language models,'' in \emph{IEEE International Conference on Acoustics, Speech and Signal Processing}.\hskip 1em plus 0.5em minus 0.4em\relax IEEE, 2024, pp. 4625--4629.

\bibitem{qu2024provably}
W.~Qu, D.~Yin, Z.~He, W.~Zou, T.~Tao, J.~Jia, and J.~Zhang, ``Provably robust multi-bit watermarking for ai-generated text via error correction code,'' \emph{arXiv preprint arXiv:2401.16820}, 2024.

\bibitem{kim2024wouaf}
C.~Kim, K.~Min, M.~Patel, S.~Cheng, and Y.~Yang, ``Wouaf: Weight modulation for user attribution and fingerprinting in text-to-image diffusion models,'' in \emph{Proceedings of the IEEE/CVF Conference on Computer Vision and Pattern Recognition}, 2024, pp. 8974--8983.

\bibitem{feng2023catch}
W.~Feng, J.~He, J.~Zhang, T.~Zhang, W.~Zhou, W.~Zhang, and N.~Yu, ``Catch you everything everywhere: Guarding textual inversion via concept watermarking,'' \emph{arXiv preprint arXiv:2309.05940}, 2023.

\bibitem{liu2024groot}
W.~Liu, Y.~Li, D.~Lin, H.~Tian, and H.~Li, ``Groot: Generating robust watermark for diffusion-model-based audio synthesis,'' in \emph{ACM Multimedia}, 2024.

\bibitem{li2022untargeted}
Y.~Li, Y.~Bai, Y.~Jiang, Y.~Yang, S.-T. Xia, and B.~Li, ``Untargeted backdoor watermark: Towards harmless and stealthy dataset copyright protection,'' \emph{Advances in Neural Information Processing Systems}, vol.~35, pp. 13\,238--13\,250, 2022.

\bibitem{guo2023domain}
J.~Guo, Y.~Li, L.~Wang, S.-T. Xia, H.~Huang, C.~Liu, and B.~Li, ``Domain watermark: Effective and harmless dataset copyright protection is closed at hand,'' in \emph{Annual Conference on Neural Information Processing Systems}, 2023.

\bibitem{li2023black}
Y.~Li, M.~Zhu, X.~Yang, Y.~Jiang, T.~Wei, and S.-T. Xia, ``Black-box dataset ownership verification via backdoor watermarking,'' \emph{IEEE Transactions on Information Forensics and Security}, vol.~18, pp. 2318--2332, 2023.

\bibitem{wei2024pointncbw}
C.~Wei, Y.~Wang, K.~Gao, S.~Shao, Y.~Li, Z.~Wang, and Z.~Qin, ``Pointncbw: Towards dataset ownership verification for point clouds via negative clean-label backdoor watermark,'' \emph{IEEE Transactions on Information Forensics and Security}, 2024.

\bibitem{guo2024zero}
J.~Guo, Y.~Li, R.~Chen, Y.~Wu, C.~Liu, and H.~Huang, ``Zeromark: Towards dataset ownership verification without disclosing watermarks,'' in \emph{Annual Conference on Neural Information Processing Systems}, 2024.

\bibitem{he2024difficulty}
Y.~He, B.~Li, Y.~Wang, M.~Yang, J.~Wang, H.~Hu, and X.~Zhao, ``Is difficulty calibration all we need? towards more practical membership inference attacks,'' in \emph{ACM SIGSAC Conference on Computer and Communications Security}, 2024.

\bibitem{ye2024duaw}
X.~Ye, H.~Huang, J.~An, and Y.~Wang, ``Duaw: Data-free universal adversarial watermark against stable diffusion customization,'' in \emph{ICLR 2024 Workshop on Secure and Trustworthy Large Language Models}, 2024.

\bibitem{cui2023diffusionshield}
Y.~Cui, J.~Ren, H.~Xu, P.~He, H.~Liu, L.~Sun, Y.~Xing, and J.~Tang, ``Diffusionshield: A watermark for copyright protection against generative diffusion models,'' \emph{arXiv preprint arXiv:2306.04642}, 2023.

\bibitem{zhu2024watermark}
P.~Zhu, T.~Takahashi, and H.~Kataoka, ``Watermark-embedded adversarial examples for copyright protection against diffusion models,'' in \emph{Proceedings of the IEEE/CVF Conference on Computer Vision and Pattern Recognition}, 2024, pp. 24\,420--24\,430.

\bibitem{li2025towards}
B.~Li, Y.~Wei, Y.~Fu, Z.~Wang, Y.~Li, J.~Zhang, R.~Wang, and T.~Zhang, ``Towards reliable verification of unauthorized data usage in personalized text-to-image diffusion models,'' in \emph{IEEE Symposium on Security and Privacy}, 2025, pp. 73--73.

\bibitem{lau2024protecting}
G.~K.~R. Lau, X.~Niu, H.~Dao, J.~Chen, C.-S. Foo, and B.~K.~H. Low, ``Protecting text ip in the era of llms with robust and scalable watermarking,'' in \emph{ICML Workshop on Generative AI+ Law}, 2024.

\bibitem{adi2018turning}
Y.~Adi, C.~Baum, M.~Cisse, B.~Pinkas, and J.~Keshet, ``Turning your weakness into a strength: Watermarking deep neural networks by backdooring,'' in \emph{USENIX Security}, 2018.

\bibitem{shao2025explanation}
S.~Shao, Y.~Li, H.~Yao, Y.~He, Z.~Qin, and K.~Ren, ``Explanation as a watermark: Towards harmless and multi-bit model ownership verification via watermarking feature attribution,'' in \emph{Network and Distributed System Security Symposium}, 2025.

\bibitem{li2022defending}
Y.~Li, L.~Zhu, X.~Jia, Y.~Jiang, S.-T. Xia, and X.~Cao, ``Defending against model stealing via verifying embedded external features,'' in \emph{AAAI conference on artificial intelligence}, 2022.

\bibitem{yang2023watermarking}
W.~Yang, S.~Shao, Y.~Yang, X.~Liu, X.~Liu, Z.~Xia, G.~Schaefer, and H.~Fang, ``Watermarking in secure federated learning: A verification framework based on client-side backdooring,'' \emph{ACM Transactions on Intelligent Systems and Technology}, vol.~15, no.~1, pp. 1--25, 2023.

\bibitem{shao2024fedtracker}
S.~Shao, W.~Yang, H.~Gu, Z.~Qin, L.~Fan, and Q.~Yang, ``Fedtracker: Furnishing ownership verification and traceability for federated learning model,'' \emph{IEEE Transactions on Dependable and Secure Computing}, 2024.

\bibitem{wu2020watermarking}
H.~Wu, G.~Liu, Y.~Yao, and X.~Zhang, ``Watermarking neural networks with watermarked images,'' \emph{IEEE Transactions on Circuits and Systems for Video Technology}, vol.~31, no.~7, pp. 2591--2601, 2020.

\bibitem{zhang2020model}
J.~Zhang, D.~Chen, J.~Liao, H.~Fang, W.~Zhang, W.~Zhou, H.~Cui, and N.~Yu, ``Model watermarking for image processing networks,'' in \emph{Proceedings of the AAAI conference on artificial intelligence}, vol.~34, no.~07, 2020, pp. 12\,805--12\,812.

\bibitem{peng2023intellectual}
S.~Peng, Y.~Chen, C.~Wang, and X.~Jia, ``Intellectual property protection of diffusion models via the watermark diffusion process,'' \emph{arXiv preprint arXiv:2306.03436}, 2023.

\bibitem{zhang2024suppressing}
L.~Zhang, Y.~Liu, X.~Zhang, and H.~Wu, ``Suppressing high-frequency artifacts for generative model watermarking by anti-aliasing,'' in \emph{Proceedings of the 2024 ACM Workshop on Information Hiding and Multimedia Security}, 2024, pp. 223--234.

\bibitem{liu2023watermarking}
Y.~Liu, Z.~Li, M.~Backes, Y.~Shen, and Y.~Zhang, ``Watermarking diffusion model,'' \emph{arXiv preprint arXiv:2305.12502}, 2023.

\bibitem{fang2017generating}
T.~Fang, M.~Jaggi, and K.~Argyraki, ``Generating steganographic text with lstms,'' \emph{arXiv preprint arXiv:1705.10742}, 2017.

\bibitem{yang2018rnn}
Z.-L. Yang, X.-Q. Guo, Z.-M. Chen, Y.-F. Huang, and Y.-J. Zhang, ``Rnn-stega: Linguistic steganography based on recurrent neural networks,'' \emph{IEEE Transactions on Information Forensics and Security}, vol.~14, no.~5, pp. 1280--1295, 2018.

\bibitem{liu2024generative}
Y.~Liu, C.~Wang, J.~Wang, B.~Ou, and X.~Liao, ``Generative steganography via live comments on streaming video frames,'' \emph{IEEE Transactions on Computational Social Systems}, 2024.

\bibitem{wu2024generative}
J.~Wu, Z.~Wu, Y.~Xue, J.~Wen, and W.~Peng, ``Generative text steganography with large language model,'' in \emph{Proceedings of the 32nd ACM International Conference on Multimedia}, 2024, pp. 10\,345--10\,353.

\bibitem{cao2022generative}
Y.~Cao, Z.~Zhou, C.~Chakraborty, M.~Wang, Q.~J. Wu, X.~Sun, and K.~Yu, ``Generative steganography based on long readable text generation,'' \emph{IEEE transactions on computational social systems}, 2022.

\bibitem{hayes2017generating}
J.~Hayes and G.~Danezis, ``Generating steganographic images via adversarial training,'' \emph{Advances in neural information processing systems}, vol.~30, 2017.

\bibitem{zhang2019steganogan}
K.~A. Zhang, A.~Cuesta-Infante, L.~Xu, and K.~Veeramachaneni, ``Steganogan: High capacity image steganography with gans,'' \emph{arXiv preprint arXiv:1901.03892}, 2019.

\bibitem{zhang2023steganography}
J.~Zhang, K.~Chen, W.~Li, W.~Zhang, and N.~Yu, ``Steganography with generated images: Leveraging volatility to enhance security,'' \emph{IEEE Transactions on Dependable and Secure Computing}, 2023.

\bibitem{jiang2020smartsteganogaphy}
S.~Jiang, D.~Ye, J.~Huang, Y.~Shang, and Z.~Zheng, ``Smartsteganogaphy: Light-weight generative audio steganography model for smart embedding application,'' \emph{Journal of Network and Computer Applications}, vol. 165, p. 102689, 2020.

\bibitem{chen2021learning}
L.~Chen, R.~Wang, D.~Yan, and J.~Wang, ``Learning to generate steganographic cover for audio steganography using gan,'' \emph{IEEE Access}, vol.~9, pp. 88\,098--88\,107, 2021.

\bibitem{li2023coverless}
J.~Li, K.~Wang, and X.~Jia, ``A coverless audio steganography based on generative adversarial networks,'' \emph{Electronics}, vol.~12, no.~5, p. 1253, 2023.

\bibitem{mao2024covert}
X.~Mao, X.~Hu, W.~Peng, Z.~Gan, Z.~Qian, X.~Zhang, and S.~Li, ``From covert hiding to visual editing: robust generative video steganography,'' in \emph{Proceedings of the 32nd ACM International Conference on Multimedia}, 2024, pp. 2757--2765.

\bibitem{huang2022cmua}
H.~Huang, Y.~Wang, Z.~Chen, Y.~Zhang, Y.~Li, Z.~Tang, W.~Chu, J.~Chen, W.~Lin, and K.-K. Ma, ``Cmua-watermark: A cross-model universal adversarial watermark for combating deepfakes,'' in \emph{Proceedings of the AAAI Conference on Artificial Intelligence}, vol.~36, no.~1, 2022, pp. 989--997.

\bibitem{tan2023somewhat}
M.~Tan, T.~Wang, and S.~Jha, ``A somewhat robust image watermark against diffusion-based editing models,'' \emph{arXiv preprint arXiv:2311.13713}, 2023.

\bibitem{xu2022robust}
Y.~Xu, C.~Mou, Y.~Hu, J.~Xie, and J.~Zhang, ``Robust invertible image steganography,'' in \emph{Proceedings of the IEEE/CVF conference on computer vision and pattern recognition}, 2022, pp. 7875--7884.

\bibitem{zhao2023invisible}
X.~Zhao, K.~Zhang, Z.~Su, S.~Vasan, I.~Grishchenko, C.~Kruegel, G.~Vigna, Y.-X. Wang, and L.~Li, ``Invisible image watermarks are provably removable using generative ai,'' \emph{arXiv preprint arXiv:2306.01953}, 2023.

\bibitem{pan2024finding}
M.~Pan, Z.~Wang, X.~Dong, V.~Sehwag, L.~Lyu, and X.~Lin, ``Finding needles in a haystack: A black-box approach to invisible watermark detection,'' \emph{arXiv preprint arXiv:2403.15955}, 2024.

\bibitem{tan2014stacked}
S.~Tan and B.~Li, ``Stacked convolutional auto-encoders for steganalysis of digital images,'' in \emph{Signal and information processing association annual summit and conference (APSIPA), 2014 Asia-Pacific}, 2014, pp. 1--4.

\bibitem{wang2021watermark}
R.~Wang, C.~Lin, Q.~Zhao, and F.~Zhu, ``Watermark faker: towards forgery of digital image watermarking,'' in \emph{2021 IEEE International Conference on Multimedia and Expo (ICME)}, 2021, pp. 1--6.

\bibitem{zhao2023generative}
X.~Zhao, K.~Zhang, Y.-X. Wang, and L.~Li, ``Generative autoencoders as watermark attackers: Analyses of vulnerabilities and threats,'' in \emph{ICLR 2023}, 2023.

\bibitem{he2024can}
Z.~He, B.~Zhou, H.~Hao, A.~Liu, X.~Wang, Z.~Tu, Z.~Zhang, and R.~Wang, ``Can watermarks survive translation? on the cross-lingual consistency of text watermark for large language models,'' \emph{arXiv preprint arXiv:2402.14007}, 2024.

\bibitem{wu2024bypassing}
Q.~Wu and V.~Chandrasekaran, ``Bypassing llm watermarks with color-aware substitutions,'' \emph{arXiv preprint arXiv:2403.14719}, 2024.

\bibitem{kassis2024unmarker}
A.~Kassis and U.~Hengartner, ``Unmarker: A universal attack on defensive watermarking,'' \emph{arXiv preprint arXiv:2405.08363}, 2024.

\bibitem{chang2024watermark}
H.~Chang, H.~Hassani, and R.~Shokri, ``Watermark smoothing attacks against language models,'' \emph{arXiv preprint arXiv:2407.14206}, 2024.

\bibitem{zhang2024large}
Z.~Zhang, X.~Zhang, Y.~Zhang, L.~Y. Zhang, C.~Chen, S.~Hu, A.~Gill, and S.~Pan, ``Large language model watermark stealing with mixed integer programming,'' \emph{arXiv preprint arXiv:2405.19677}, 2024.

\bibitem{fernandez2022watermarking}
P.~Fernandez, A.~Sablayrolles, T.~Furon, H.~J{\'e}gou, and M.~Douze, ``Watermarking images in self-supervised latent spaces,'' in \emph{ICASSP 2022-2022 IEEE International Conference on Acoustics, Speech and Signal Processing (ICASSP)}.\hskip 1em plus 0.5em minus 0.4em\relax IEEE, 2022, pp. 3054--3058.

\bibitem{tancik2020stegastamp}
M.~Tancik, B.~Mildenhall, and R.~Ng, ``Stegastamp: Invisible hyperlinks in physical photographs,'' in \emph{Proceedings of the IEEE/CVF conference on computer vision and pattern recognition}, 2020, pp. 2117--2126.

\bibitem{blueprint}
T.~W. House, ``Blueprint for an ai bill of rights,'' 2022, https://www.whitehouse.gov/ostp/ai-bill-of-rights/.

\bibitem{riskframework}
N.~I. of~Standards and Technology, ``Ai risk management framework,'' 2024, https://www.nist.gov/itl/ai-risk-management-framework.

\bibitem{MaliciousAct}
U.~S. Congress, ``Malicious deep fake prohibition act,'' 2018, https://www.congress.gov/bill/115th-congress/senate-bill/3805/related-bills.

\bibitem{euact}
{The European Union}, ``Eu artificial intelligence act,'' 2024, \newline\url{https://artificialintelligenceact.eu/}.

\bibitem{gccontentregulation}
{Cyberspace Administration of China}, ``Administrative provisions on deep synthesis in internet-based information services,'' 2022, \newline\url{https://www.gov.cn/zhengce/zhengceku/2022-12/12/content_5731431.htm}.

\bibitem{draft}
------, ``Administrative measures for generative artificial intelligence services,'' 2023, \newline\url{https://www.cac.gov.cn/2023-07/13/c_1690898327029107.htm}.

\bibitem{GBdraft}
------, ``Cybersecurity technology - labeling method for content generated by artificial intelligence(draft for public consultation),'' 2024, \newline\url{https://www.cac.gov.cn/2024-09/14/c_1728000676244628.htm}.

\bibitem{CybersecurityStandards}
------, ``Cybersecurity standards—content labeling methods for generative artificial intelligence services,'' 2023, \newline\url{https://www.tc260.org.cn/upload/2023-08-25/1692961404507050376.pdf}.

\bibitem{GB}
------, ``Cybersecurity technology - labeling method for content generated by artificial intelligence,'' 2024, \newline\url{https://www.tc260.org.cn/upload/2024-09-14/1726290836419027596.pdf}.

\bibitem{GDPR}
{European Parliament}, ``General data protection regulation,'' 2016, \newline\url{https://eur-lex.europa.eu/eli/reg/2022/2065/oj}.

\bibitem{Directive}
{THE EUROPEAN PARLIAMENT}, ``Directive,'' 2016, \newline\url{https://eur-lex.europa.eu/legal-content/EN/TXT/?uri=CELEX:32016L0943}.

\bibitem{PDPA}
{Singapore}, ``Personal data protection regulations,'' 2021, \newline\url{https://sso.agc.gov.sg/SL-Supp/S63-2021/Published/20210129?DocDate=20210129}.

\bibitem{PIPL}
{The Standing Committee of the National People's Congress of China}, ``Personal information protection law of the people's republic of china,'' 2021, \newline\url{https://www.gov.cn/xinwen/2021-08/20/content_5632486.htm}.

\bibitem{doi:10.1126/science.adp1848}
\BIBentryALTinterwordspacing
R.~Bommasani, S.~Kapoor, K.~Klyman, S.~Longpre, A.~Ramaswami, D.~Zhang, M.~Schaake, D.~E. Ho, A.~Narayanan, and P.~Liang, ``Considerations for governing open foundation models,'' \emph{Science}, vol. 386, no. 6718, pp. 151--153, 2024. [Online]. Available: \url{https://www.science.org/doi/abs/10.1126/science.adp1848}
\BIBentrySTDinterwordspacing

\bibitem{Digitalservicesact}
{The European Union}, ``Digital services act,'' 2022, \newline\url{https://eur-lex.europa.eu/eli/reg/2022/2065/oj}.

\bibitem{Ecological}
{Cyberspace Administration of China}, ``Regulations on ecological governance of network information of the people's republic of china,'' 2019, https://www.gov.cn/zhengce/zhengceku/2020-11/25/content\_5564110.htm.

\end{thebibliography}

\end{document}